\documentclass[]{elsarticle}

\usepackage{lineno,hyperref}
\usepackage{enumitem}
\modulolinenumbers[5]

\journal{C. R. Physique}

\begin{document}

\begin{frontmatter}

    \title{The unit of time: present and future directions}

    \author{S\'ebastien Bize}
    \address{SYRTE, Observatoire de Paris, Universit\'e PSL, CNRS, Sorbonne Universit\'e, LNE, \\ 61 avenue de l'Observatoire, 75014 Paris, France}

%
%

    \begin{abstract}
    Some 50~years ago, physicists, and after them the entire world, started to found their time reference on atomic properties instead of motions of the Earth that have been in use since the origin. Far from being an arrival point, this decision marked the beginning of an adventure characterized by a 6 orders of magnitude improvement in the uncertainty of realization of atomic frequency and time references. Ever progressing atomic frequency standards and time references derived from them are key resources for science and for society. We will describe how the unit of time is realized with a fractional accuracy approaching $10^{-16}$ and how it is delivered to users via the elaboration of the international atomic time. We will describe the tremendous progress of optical frequency metrology over the last 20~years which led to a novel generation of optical frequency standards with fractional uncertainties of $10^{-18}$. We will describe work toward a possible redefinition of the SI second based on such standards. We will describe existing and emerging applications of atomic frequency standards in science.
    \end{abstract}



\end{frontmatter}


\section{The unit of time}\label{sec_unit_of_time}

In 1967, the 13$^{\mathrm{th}}$ General Conference on Weights and Measures (CGPM) defined the SI second as \emph{``the duration of 9~192~631~770 periods of the radiation corresponding to the transition between two hyperfine levels of the ground state of the cesium 133 atom''} \cite{CGPM1967}. Atoms have quantized energy levels. To any pair of quantized levels of energy $E_g$ and $E_e$, a frequency $\nu$ is associated via Planck-Einstein's relation: $h\nu = E_e - E_g$. One fundamental idea behind the definition is that these atomic frequencies are perfectly stable and universal. Observations support this idea. Atoms can be regarded as ``perfect'' frequency standards given by nature. The immutability of atomic frequencies is integrated into fundamental theories underpinning physics: general relativity and the standard model of particle physics.

\begin{figure}[h]
    \centering
    \includegraphics[width=\columnwidth]{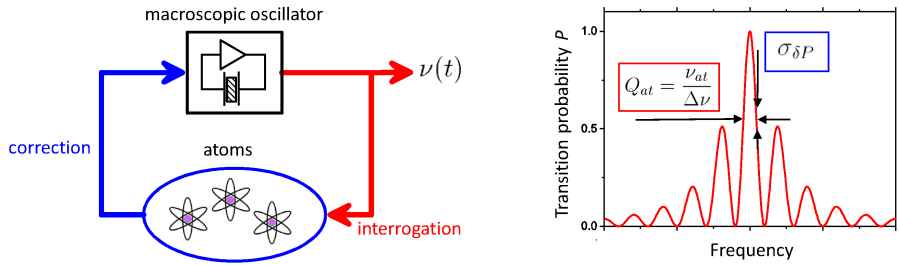}
\caption{On the left: architecture of highly accurate atomic frequency standards. An oscillator generates a macroscopic, practically usable electromagnetic signal. A fraction of this signal is split to probe the atomic transition chosen as reference. The response of the atomic sample is detected to determine the probability for exciting the transition. This information is used to stabilize the frequency of the oscillator to the spectroscopic signal. On the right: representative spectroscopic signal. Key features of this signal are the atomic quality factor $Q_{at}=\nu_{at}/\Delta\nu$ and the noise in the measurement of the transition probability $\sigma_{\delta P}$. The frequency $\nu(t)$ of the usable signal does not coincide exactly with the unperturbed atomic frequency $\nu_{at}$: $\nu(t)=\nu_{at}\times (1 + \varepsilon + y(t))$ where $\varepsilon$ is a fractional frequency offset and $y(t)$ represents fractional frequency fluctuations.}
\label{fig_architecture}
\end{figure}

Accessing the atomic frequency and transferring its qualities to a macroscopic usable signal requires a device called atomic frequency standard. The typical architecture of such device is shown and explained in fig.~\ref{fig_architecture}. The output frequency $\nu(t)$ does not coincide with the atomic frequency $\nu_{at}$. It is perturbed by noises and biases of both technical and fundamental nature. These noises and biases determines two mains characteristics of a given atomic frequency standard. Noises limit the uncertainty with which a frequency measurement can be made with this standard in a given duration. It is characterized by the fractional frequency instability $\sigma_y(\tau)$ which is a function of the measurement duration $\tau$. Biases offset the mean output frequency with respect to the unperturbed atomic frequency $\nu_{at}$. If known and stable, a bias can be taken into account. What really matters is the uncertainty to which biases are known. This uncertainty, often reported in fractional terms, defines the level to which the standard actually gives access to the unperturbed atomic frequency. It summarizes the capability of the standard to realize the SI second (for $^{133}$Cs primary standards), to be used for fundamental physics and for other applications.

In practice, only well-chosen atomic transitions are suitable to realize standards with lowest stability and uncertainty. Levels must have long lifetimes to enable high atomic quality factor $Q_{at}$ (see fig.~\ref{fig_architecture}). Transition must have low sensitivity to external fields (electric, magnetic, thermal radiation, etc.). Atomic structure must be compatible with methods needed to manipulate and detect atoms. It is also essential to consider practical criteria such as reliability, operability, possibility to generate and use the interrogation field at the transition frequency. Given the state of knowledge and technology at the time of the 13$^{\mathrm{th}}$ CGPM, a frequency standard based on the $^{133}$Cs ground state hyperfine transition was one of the best possibilities, almost 2 decades after the first observation of the transition \cite{Lyons1952}. It had shown sufficient maturity and had been accurately measured with respect to the ephemeris time \cite{Essen1955}\cite{Markowitz1958}. Since then, fundamental and technological breakthroughs in many areas lead to major changes in ways to realize and use highly accurate atomic frequency standards. Their uncertainty improved by 6 orders of magnitudes. References \cite{Quinn2005} and \cite{Salomon2015} provide an overview of these developments.

In November 2018, the 26$^{\mathrm{th}}$ CGPM adopted a major redefinition of the international system of units. The redefinition concerns the units of mass, of temperature, of electrical current and of amount of substance. The new system is defined by adopting conventional values for Planck's constant $h$, for Boltzmann's constant $k_B$, for the elementary charge $e$ and for Avogadro's constant $N_A$, similarly to what is already done since 1983 \cite{CGPM1983} with the speed of light $c$ to define the unit of length. In essence, the definition of the second remains unchanged. The form of the definition however is modified to resemble those of other units, i.e. \emph{``the International System of Units, the SI, is the system of units in which: the unperturbed ground state hyperfine transition frequency of the caesium 133 atom $\Delta\nu_{\mathrm{Cs}}$ is 9~192~631~770~Hz, [...]''} \cite{CGPM2018}.

\section{Research on highly accurate atomic frequency standards}

Previous section~\ref{sec_unit_of_time} gave fundamental concepts behind atomic time and atomic frequency standards. These concepts are amazingly simple and used for more than 5 decades. Still, they continue being at the basis of an active field of research in the following ways.

Search for advancing atomic frequency standards with extremely low uncertainties goes hand in hand with exploration of atomic systems interacting with electromagnetic fields. Progress in atomic frequency standards reveals new phenomena and provides means of investigating them. And vice versa, progress in other areas (e.g. laser-cooling of atoms, physics of collisions and interactions, quantum entanglement,...) enables improvement of atomic frequency standards.

These devices with extremely low uncertainties are tools of choice to probe the structure of space-time and to test fundamental laws of nature. They provide experimental inputs to the quest for a unified theory of gravitation and quantum mechanics \cite{Uzan2011}\cite{Will2014}\cite{Safronova2018} and to the search for dark matter \cite{VanTilburg2015a}\cite{Hees2016} and to other applications.

On the applied side, atomic frequency standards are key to major applications of utmost importance in modern science and society. The practical realization of the unit of time of the SI system is one of them, in particular via the elaboration of the international atomic time and of the universal coordinate time (TAI/UTC). Another example is global navigation satellite systems (GNSS) such as GPS, GALILEO, etc. In turn, these realizations are key for telecommunications, transports, finance, digital economy, etc.

Search for ultra-low uncertainties in atomic frequency standards is a steady driver for innovation in multiple technological areas such as lasers, low noise electronics, ultra-stable oscillators. Also, it creates knowledge necessary to define trade-offs between performance and other requirements for industrial frequency standards or space clocks. This knowledge is often applicable not only to novel frequency standards but also to other types of atom-based instruments like, for example, accelerometers and gyrometers based on matter wave interferometry or magnetometers.

Progress in reducing the uncertainty of atomic frequency standards leads to novel applications. For instance, because of its much lower uncertainty, the new generation of optical frequency standards enable chronometric geodesy, i.e. the determination Earth gravitational potential differences via the measurement of Einstein's gravitational red shift.

\section{Primary frequency standards based on atomic fountains}

\begin{figure}[h]
    \centering
    \includegraphics[width=\columnwidth]{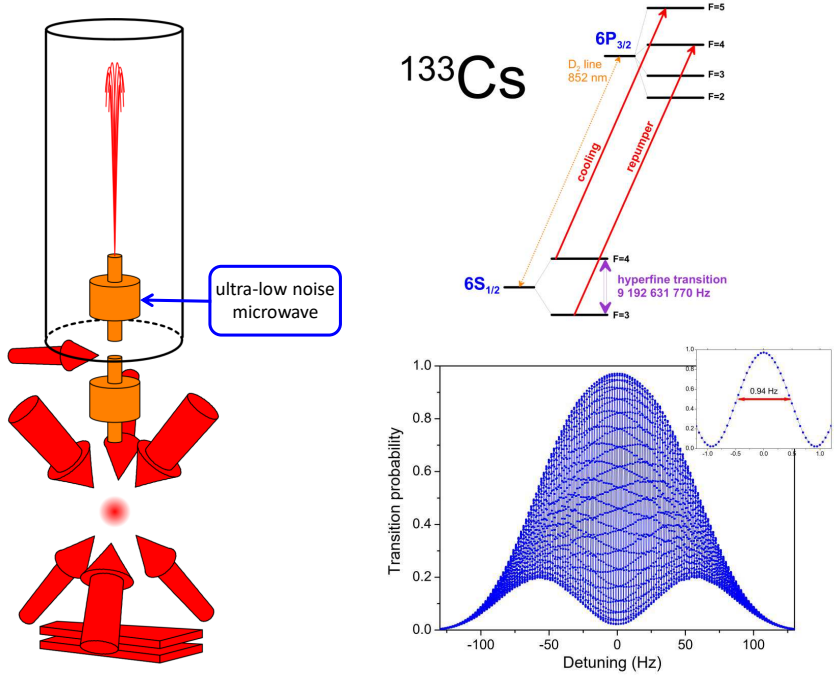}
\caption{On the left: Schematic of an atomic fountain frequency standard. A cloud of cold atoms is captured at the crossing of 3 pairs of counter-propagating laser beams. It is launched upwards at a typical speed of $4$~m.s$^{-1}$ and with temperature of $\sim 1~\mu$K. During the ballistic flight, atoms are state-selected with a first microwave interaction and a push laser beam. Then, they pass upward and downward through a microwave cavity where they interact with the signal from the interrogation oscillator. They continue falling through two laser beam sheets in the detection region. On the right: Top: Energy levels of $^{133}$Cs showing the ground-state hyperfine transition defining the SI second. In red, transitions used for laser-cooling and detection. Bottom: Spectroscopy of the cesium hyperfine transition in an atomic fountain showing Ramsey fringes with a width $<1$~Hz and an atomic quality factor $Q_{at}$ of $10^{10}$. Each point is a single measurement of the transition probability at a rate of $\sim 1$ per second with a typical noise $\sigma_{\delta P}\sim 2\times 10^{-4}$.}
\label{fig_fountain}
\end{figure}

The first generation of $^{133}$Cs primary frequency standards, which led to the adoption of the atomic time, was based on the thermal atomic beam technology and the separated oscillatory fields method \cite{Ramsey1990}\cite{Vanier1989}. The development of laser cooling of atoms in the 1980's \cite{CohenTannoudji1998}\cite{Phillips1998}\cite{Chu1998} enables a second generation called atomic fountains \cite{Kasevich1989}\cite{Clairon1991}\cite{Clairon1995}. Atomic samples laser-cooled to temperature near $1~\mu$K enable atomic quality factors higher than $10^{10}$, a factor 100 higher than in atomic beam standards. Figure~\ref{fig_fountain} displays a schematic of an atomic fountain. When an ultra-low noise microwave source is used to interrogate the atomic transition the quantum projection noise limit is reached \cite{Santarelli1999}.
This limit is the fundamental limit set by the quantum measurement process for un-entangled particles. This yields short term fractional frequency instabilities as low as $1.6\times 10^{-14}$ at $1$~s \cite{Guena2012}. Nowadays, the accuracy of best atomic fountains ranges between 1 and 3 parts in $10^{16}$. This is the result of refining models of the interrogation and detection processes, and of stringently testing these models with experiments. Such experiments typically require many frequency measurements with statistical uncertainties near $10^{-16}$, which is reached, in the very best case, after several full days of measurement duration. In other words, fountains have reached the situation where the stability imposes a practical limit to studying systematic shifts and improving the accuracy. Example of effects whose modelling improved significantly includes effects of phase gradients in the interrogation microwave cavity \cite{Guena2011}\cite{Li2010b}, effects of cold collisions \cite{Pereira2002}\cite{Papoular2012a}\cite{Szymaniec2007}, effects of collisions with background gases \cite{Gibble2013}\cite{Szymaniec2014} and effects of the microwave field on external atomic motion (``microwave lensing'') \cite{Gibble2006}\cite{Li2011}\cite{Gibble2014}. Implementation of fountains with cryogenic interrogation environment enabled a new direct measurement of effects of thermal radiation shift (``blackbody radiation shift'') \cite{Itano1982}\cite{Jefferts2014}\cite{Heavner2014}\cite{Levi2014}.
Recently, a complete accuracy budget was published for continuous fountain at the level of $1.99\times 10^{-15}$ \cite{Jallageas2018}. More details on atomic fountains can be found for instance in \cite{Wynands2005},
\cite{Guena2012} and \cite{Riehle2016}.

The $^{87}$Rb ground state hyperfine transition is also used to realise atomic fountain frequency standards. LNE-SYRTE developed a dual fountain using Rb and Cs which truly realizes two state-of-the-art microwave standards in a single setup \cite{Guena2010}\cite{Guena2014}. The Rb part of this fountain has an uncertainty of $3.2\times 10^{-16}$ similar to the one of best cesium fountains. This dual fountain enables highly accurate comparisons of the $^{87}$Rb and $^{133}$Cs hyperfine frequencies.  These comparisons find many applications (see sections~\ref{sec_fundamental} and \ref{sec_redefinition} below). In particular, they led to the adoption of this transition has a secondary representation of the SI second \cite{BIPM_LOR}\cite{Guena2014}.

Agreement between fountains is well tested by the means of specific remote comparisons by satellite methods \cite{Bauch2006} and recently by optical links between LNE-SYRTE and PTB \cite{Guena2017}. Also, the elaboration of TAI provides a vehicle to compare fountain frequency standards (see section~\ref{sec_TAI} below).

\section{Elaboration of TAI: accuracy of atomic fountains delivered to users}\label{sec_TAI}

One essential outcome of time and frequency metrology is the construction of the international atomic time (TAI) and of the universal coordinate time (UTC). It is worth noting here that the 26$^{\mathrm{th}}$ CGPM adopted a resolution on the definition of time scales \cite{CGPM2018}, which corrects for the lack so far of a self-contained definition of TAI. We remind that a meaningful definition and realization of a time scale valid globally in the vicinity of the Earth requires the framework of general relativity, in particular to properly account for Einstein's gravitational redshift which is about $10^{-16}$ per meter of elevation at the surface of the Earth. A description of elaboration of TAI by the BIPM can be found for example in \cite{Petit2015a} and references therein. By means of satellite-based comparisons, data from about 450 continuously operated commercial clocks are used to compute the free atomic time scale (EAL). The large number of devices and locations ensure the permanence of EAL. Data from a much smaller number of primary and secondary frequency standards (about 20) in an even more limited number of institutes are used to calibrate the frequency of EAL against the $^{133}$Cs hyperfine transition, according to the definition of the second, and to steer EAL to realize TAI. UTC is derived from TAI by inserting leap seconds that maintain agreement with universal time (UT1) and the observed rotation of the Earth. The result of this process is published by the BIPM in its monthly \emph{Circular~T} \cite{CircularT}. This information, in turn, is used by national metrology institutes and other participants to steer their local physical representation of UTC from which disseminations in society start by various means.

\begin{figure}[h]
    \centering
    \includegraphics[width=0.8\columnwidth]{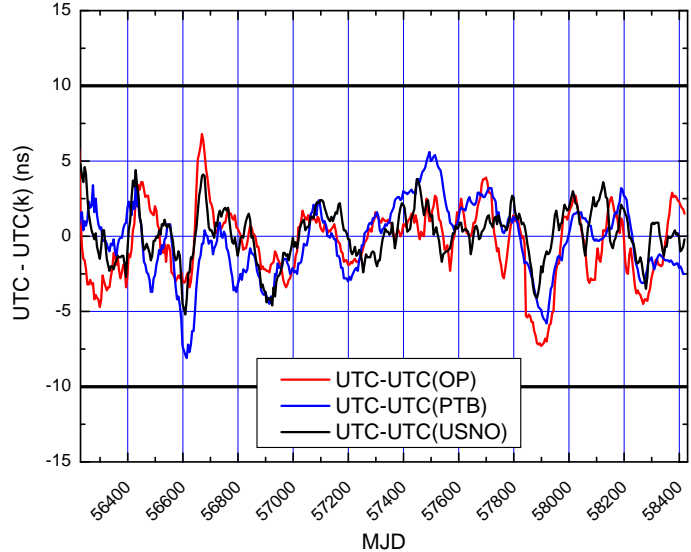}
\caption{Local realizations of UTC using atomic fountain data. The plot shows time differences UTC-UTC(k) as a function of the Modified Julian Date (MJD) for UTC(USNO), UTC(PTB) and UTC(OP). Data are shown between October 29$^{\mathrm{th}}$ 2012 and November 08$^{\mathrm{th}}$ 2018.}
\label{fig_timescales}
\end{figure}

\emph{Circular~T} and calculations done at the BIPM also yield an estimation of the performance of TAI and provide a vehicle to compare frequency standards worldwide. Differences between frequency standards observed by this means are consistent with uncertainties \cite{Parker2012}\cite{Petit2013}. The accuracy to which the scale interval is determined with respect to $^{133}$Cs hyperfine transition now reaches $2\times 10^{-16}$ \cite{Petit2015a}. In other words, performance of frequency standards is transferred to the time scale and thereby to users. The $2\times 10^{-16}$ fractional frequency uncertainty translates into an error of less than $10$~ns after one year. This improvement by a factor of 10 since 2000 is the result of the commitment of a few metrology institutes to provide regular calibrations of TAI. Nowadays, about 4 or 5 calibrations by fountains are typically available each month, as can be seen in section~3 of \emph{Circular~T} \cite{CircularT}. Over the last 15 years, LNE-SYRTE made 40\% of all worldwide calibrations with fountains. The adoption of secondary representations of the SI second (see \cite{BIPM_LOR} and section~\ref{sec_redefinition} below) led to the possibility to calibrate TAI with other atomic transitions than the $^{133}$Cs hyperfine transition. This was done for the first time with the $^{87}$Rb hyperfine transition by LNE-SYRTE which provided close to 100 calibrations by this mean \cite{Guena2014}. LNE-SYRTE also pioneered providing calibrations based on an optical transition. This was done with the $^{87}$Sr $^1$S$_0$-$^3$P$_0$ transition at 698~nm. Regular calibrations of TAI by optical frequency standards are an important prerequisite for a possible redefinition of the second based on optical transition(s) \cite{Riehle2018}. The transfer of long term stability and accuracy of primary frequency standards to TAI enables highly accurate SI-traceable frequency measurements without a local primary frequency \cite{Hachisu2017}\cite{McGrew2018a} and fundamental physics tests \cite{Ashby2007}\cite{Ashby2018}.

Progress in the reliability of atomic fountains made possible for a few institutes to steer their local realizations of UTC with fountain data \cite{Bauch2012}\cite{Peil2014}\cite{Rovera2016}. Local realization of UTC by institute k is denoted UTC(k). Figure~\ref{fig_timescales} shows time differences between such timescales and UTC. They are realized using hydrogen masers whose frequency is calibrated and steered with atomic fountains, and for the long term, with the time difference UTC(k)-UTC provided every month in \emph{Circular~T}. They typically deviate from UTC and from each others by no more than a few nanoseconds. It is worth noting another significant evolution in timekeeping, namely the rapid realization of UTC by the BIPM \cite{Petit2014}. UTCr implements faster exchange of data than \emph{Circular~T} does between participating laboratories and the BIPM. It improves the level to which a given laboratory can verify the synchronization of its UTC(k) to UTC. This is of particular interest and significance for laboratories where resources allocated to the local timescale are limited (i.e. limited to a few commercial Cs beam standards).

\section{A new generation of optical atomic frequency standards}

\begin{figure}[h]
    \centering
    \includegraphics[width=\columnwidth]{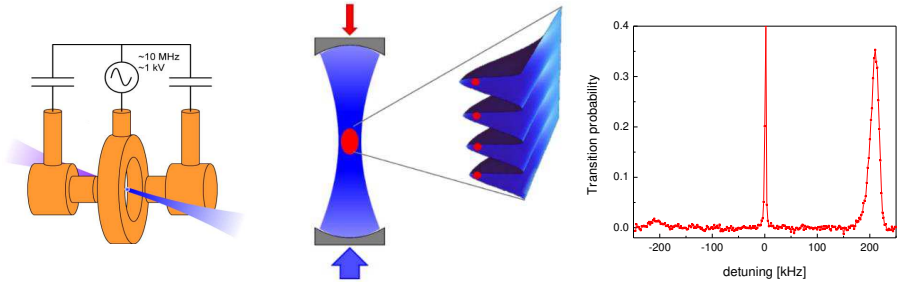}
\caption{On the left: Schematic of single ion optical frequency standard. The ion is confined in a Paul trap made of electrodes with oscillating electric potentials. The ion is laser-cooled and probed with ultra-stable laser light. Center: Schematic of optical lattice neutral atom frequency standard. An ensemble of atoms is confined in a corrugated potential formed by an intense standing laser field (blue). Atoms are probed with ultra-stable light aligned with the trap axis (red). On the right: Spectroscopy of the $^{87}$Sr reference transition at $429$~THz in a optical lattice. The spectrum is typical of Lamb-Dicke regime used in optical frequency standards. Because of confinement, the external motion of atoms has quantized vibrational levels. The spectrum exhibits resolved sidebands with frequency offsets corresponding to the trap vibrational frequency. The central ``carrier'' resonance corresponds to excitations that do not change the external atomic state. It is essentially unaffected by the first order Doppler effect and by the recoil effect \cite{Bergquist1987}. This resonance can be as narrow as few $100$~mHz corresponding to atomic quality factors $Q_{at}$ of several $10^{15}$, at the origin of the superior performance of optical frequency standards. The counterpart of using confined particles is the need to care about effects of the trapping fields. Fundamentally, the ion is held thanks to its electric charge without strong perturbation of its internal structure. It is nevertheless necessary to care about effects of micro-motions induced by the oscillating electric field \cite{Paul1990}\cite{Leibfried2003}. Trapping of neutral atoms relies on polarizing them with the intense lattice light and thereby on perturbing their internal structure. In optical frequency standards, the lattice trap can be non-perturbing if the trap wavelength has a specific wavelength called \emph{magic wavelength} where the polarizability is the same for the two levels of the reference transition \cite{Katori2001}\cite{Katori2003a}\cite{Ye2008a}.}
\label{fig_optical_standards}
\end{figure}

\begin{figure}[h]
    \centering
    \includegraphics[width=0.8\columnwidth]{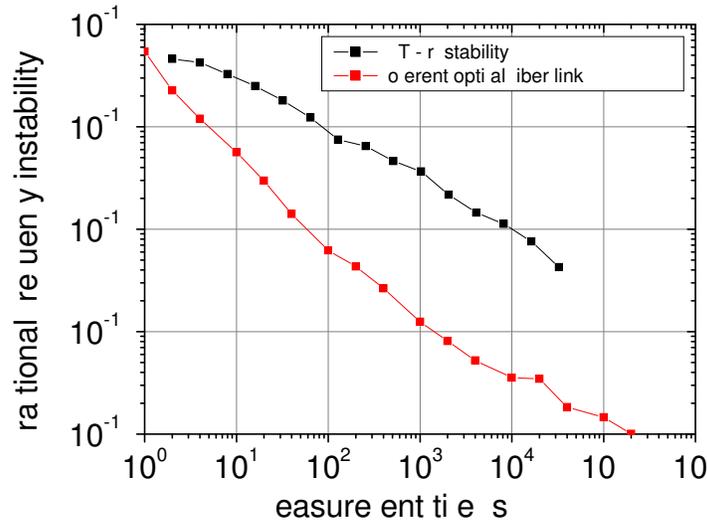}
\caption{Stability of optical frequency standard and coherent optical fiber links. In black: fraction frequency instability of a $^{87}$Sr optical lattice frequency standard from LNE-SYRTE. Curve shows the stability of a single standard (SYRTE-SrB) inferred from the analysis of a multipartite comparison with SYRTE-Sr2, NPL-Sr and PTB-Yb+. In red: stability of a Paris-Strasbourg-Paris link developed within equipex REFIMEVE+ project coordinated by LPL.}
\label{fig_stability}
\end{figure}

Optical frequency metrology is advancing at high pace, in particular since the introduction of optical frequency combs \cite{Hansch2006}\cite{Hall2006}. Optical frequency standards refer to atomic standards relying on transitions whose frequency corresponds to the optical domain of the electromagnetic spectrum. Their frequency is $10^{4}$ to $10^{5}$ higher than the $^{133}$Cs hyperfine frequency and their potential atomic quality factor can exceed $10^{16}$ instead of $10^{10}$ for atomic fountain standards. To date, both neutral atoms and ions are studied with the aim to obtain the lowest possible uncertainties. Figure~\ref{fig_optical_standards} explains the principle of operation of these two types of standards. Optical frequency standards achieve fractional frequency uncertainties close to $10^{-18}$ \cite{McGrew2018b}\cite{Takano2016}\cite{Huntemann2016}\cite{Nicholson2015}\cite{Ushijima2015}\cite{Bloom2014}\cite{Beloy2014a}\cite{Dube2014}.
For example, reference \cite{McGrew2018b} reports an uncertainty of $1.4\times 10^{-18}$ for an $^{171}$Yb optical lattice standard. References \cite{Nicholson2015} and \cite{Takano2016} report uncertainties of $2.1\times 10^{-18}$ and $4.8\times 10^{-18}$ respectively for $^{87}$Sr optical lattice standards. Reference \cite{Huntemann2016} reports an uncertainty of $3.2\times 10^{-18}$ for a $^{171}$Yb$^+$ ion standard.
Many physical effects had to be understood and controlled at an even lower level individually. Some notable ones are frequency shifts due to blackbody radiation
\cite{Middelmann2012a}\cite{Sherman2012}\cite{Nicholson2015}\cite{Bloom2014}\cite{Ushijima2015}\cite{Beloy2014a}\cite{Dube2014}\cite{Dolezal2015} and to electric fields \cite{Lodewyck2012}, shifts induced by the lattice light
\cite{Westergaard2011}\cite{Brown2017a}, shifts induced by interactions \cite{Campbell2017a}, light-shifts induced by probe light \cite{Yudin2010}. This list of example must not be considered as complete either in terms of effects or in terms of references. Comparisons between optical frequency standards using the same transition were performed with improving uncertainties \cite{Chou2010}\cite{LeTargat2013a}\cite{Bloom2013}\cite{Ushijima2015}\cite{Huang2016}, down to below $10^{-18}$ for the most recent work \cite{McGrew2018b}. A recent and more complete account on optical frequency standards can be found in \cite{Ludlow2015}. Figure~\ref{fig_stability} shows a recent measurement of the stability of a $^{87}$Sr optical lattice frequency standard from LNE-SYRTE.

Evaluation of uncertainties and comparisons at the $10^{-18}$ level are made possible by the excellent short term stabilities reached by optical frequency standards \cite{Schioppo2017}\cite{Hinkley2013a}\cite{Takamoto2011}\cite{Nicholson2012a}\cite{AlMasoudi2015}. Stabilities as low as $1.6\times 10^{-16}$ at 1~s can be observed for a single optical lattice frequency standard and even lower between two regions of the same atomic cloud \cite{Marti2018}. Ion-based optical frequency standards show significantly worse stabilities because they are so far using a single ion instead of several hundreds of atoms or more in neutral atom standards. Synchronized and correlated interrogation of 2 ion-based standard enables comparison with stability in the mid-$10^{-16}$ at 1~s \cite{Chou2011a}.
Such stabilities became possible only with the progress of ultra-stable lasers which remain the subject of active developments. Classical ultra-stable lasers based on Fabry-Perot prove being limited by thermal brownian noise in dielectric mirror coatings \cite{Numata2004}. Several approaches were and are still investigated to mitigate this limit. Extending cavity length can already give significant improvement \cite{Swallows2012}\cite{Hinkley2013a}\cite{Hafner2015}. Crystalline silicon cavity at cryogenic temperatures showed exquisite laser instability and laser linewidth ($4\times 10^{-17}$ and $<10$~mHz respectively) \cite{Kessler2012a}\cite{Matei2017}. Another promising method could be to use crystalline coatings that exhibit lower thermal noise \cite{Cole2013a}. Other approaches shift away from Fabry-Perot cavity. Prospects exist to use spectral hole burning in rare-earth doped ions in crystalline matrices at cryogenic temperatures \cite{Chen2011d}\cite{Thorpe2011a}\cite{Gobron2017}. Lasers using ultra-narrow atomic transitions are another proposed alternative \cite{Yu2007}\cite{Meiser2009a}\cite{Norcia2018}.

Interest in quantum technologies increased considerably in the last years because they promise major breakthroughs in computation, communication and sensing \cite{Wehner2018}\cite{Degen2017}. Research in optical frequency standards already gave striking examples of quantum enhanced metrology. Operation of the single Al$^+$ ion frequency standard relies on a quantum gate between Al$^+$ and a companion ion used for state readout \cite{Chou2010}\cite{Schmidt2005}\cite{Wineland2013}. To date, there is still a large potential for optical frequency standards to further exploit quantum metrology. Tailored quantum superposition of internal states can reduce sensitivity to external field perturbations \cite{Roos2006}. Entangled states of several ions can improve the stability below the quantum projection noise limit \cite{Leibfried2002}. In parallel, progress were made in designing traps that can support multiple-ion chains while maintaining low uncertainty due to motional effects \cite{Keller2019a}. A challenge for future ion-based standards will be to merge all methods in a single device. Neutral atom standards already use samples that comprise hundreds or thousands of atoms. Quantum non-destructive measurements performed on such samples can generate entangled states \cite{Bouchoule2002} which are metrologically useful like, for example, spin-squeezed states \cite{Wineland1992}. Proof-of-principle experiments using hyperfine transitions are reported in \cite{Windpassinger2008}\cite{Louchet-Chauvet2010} and in \cite{Hosten2016} where a specular 10-fold reduction (20~dB squeezing) below the quantum projection limit is obtained. Non-destructive detection by optical phase shift measurement in $^{87}$Sr optical lattice frequency standard is developed at LNE-SYRTE \cite{Lodewyck2009}\cite{Vallet2017a}. A promising sensitivity level of a few atoms only is achieved and the path toward quantum non-destructive regime is clarified. Optical lattice frequency standards constitute an excellent platform to harvest the benefit of both classical and quantum non-destructive detection. Classical non-destructive detection could help reducing dead times in the probing sequence and thereby limit the negative influence of laser frequency noise on stability. Atomic phase lock method \cite{Shiga2012}\cite{Kohlhaas2015} could be used to extend the interrogation duration beyond the probe laser coherence time. Quantum non-destructive detection could be used to beat the quantum projection noise limit. These schemes promise stabilities at $10^{-17}$ at 1~s or below. It will remain to investigate to which extent the non-destructive detection introduces additional sources of uncertainty. Many other schemes to generate entangled states do exist which are potentially interesting to improve frequency standards, like for instance \cite{Kruse2016}\cite{Leroux2010a}\cite{Leroux2010}. The present account must not be considered complete.

%
%
%

Optical frequency combs are of utmost importance for optical frequency standards. They enable comparisons between optical frequency references at different wavelengths. They enable conversion of optical frequency references to the microwave domain and thereby, connection with all existing time and frequency methods and infrastructures, in particular comparisons between optical and microwave standards. Optical frequency combs developed and found applications in many other different fields which are reviewed for example in \cite{Diddams2010}\cite{Newbury2011}. Here, we highlight developments directly connecting to highly accurate frequency standards. One key aspect was the development of reliable femtosecond lasers based on erbium-doped fiber technology. Combs based on this technology readily connect to the wavelength of $1.5~\mu$m used in optical fiber links (see below) and in best ultra-stable lasers to date. Such combs can transfer the stability between ultra-stable optical references with degradation no higher than $4\times 10^{-18}$ at 1~s, a level which surpass by far the stability of best ultra-stable lasers \cite{Nicolodi2014}. Optical to microwave conversion was also developed to minimize noises from all processes involved in the comb architecture \cite{Xie2017a}\cite{Baynes2015}. Reliable optical frequency comb systems can deal with multiple wavelengths simultaneously and support complex optical and microwave frequency standard comparisons and microwave generation applied to atomic fountains \cite{Ohmae2017}\cite{LeCoq2012}\cite{Millo2009a}\cite{Weyers2009a}\cite{Lipphardt2017}\cite{Lodewyck2016}\cite{Guena2017}\cite{Lisdat2016}\cite{Tyumenev2016}.

Optical fiber links made tremendous progress over the last decade. Coherent optical fiber links transmit an ultra-stable laser light at $1.5~\mu$m and are therefore directly adapted to the comparison of distant optical frequency standards. From initial proof of concept experiments \cite{Grosche2007}\cite{Jiang2008}\cite{Grosche2009}, this method was extended to continental distances \cite{Predehl2012}\cite{Lopez2012a}. It enabled comparison of optical frequency standards over continental distances with unprecedented stability and accuracy \cite{Lisdat2016}. Figure~\ref{fig_stability} shows the stability of a Paris-Strasbourg-Paris link obtained for 2 weeks of operation. This link enables comparison to $10^{-18}$ in less than 2000~s.
In France, the first industrial-grade link of this type was recently implemented and tested \cite{GuillouCamargo2018}. It proved the readiness of this technology for commercialization and applications. Coherent optical fiber links reach fractional frequency instability below $10^{-18}$ for measurement duration of $10^{4}$~s for length of several hundreds to 1500~km. Accuracy of the frequency transfer is verified to better than $10^{-19}$. Over continental distances, they surpass satellite-based methods by 3 to 4 orders of magnitude. Besides frequency transfer using a coherent optical carrier, time transfer using optical fibers is actively developed with, here also, the promise to surpass existing methods by orders of magnitude (see, for instance \cite{Frank2018}\cite{Lopez2015}\cite{Sliwczynski2013}\cite{Lopez2012b}). Relativistic effects in optical fiber links are studied from a theoretical standpoint in \cite{Gersl2015}.
In addition to comparisons of optical frequency standards and dissemination of time and frequency references, optical fiber links are used or considered for applications in chronometric geodesy (see section~\ref{sec_geodesy} below), in measurement of Earth's rotation \cite{Schiller2013a}\cite{Clivati2013}, in earthquake detection \cite{Marra2018} and in fundamental physics (see e.g.~\cite{Delva2017a}).


\section{Test of fundamental physical laws}\label{sec_fundamental}

Physics is underpinned by two fundamental theories: general relativity and the standard model of particle physics. This theoretical framework is extremely successful in describing a huge number of observations. Nonetheless, it is not free of significant problems. One difficulty is that the standard model is a quantum field theory while general relativity is not. This is giving an heterogeneous picture of the three fundamental interactions. To date, a unified theory of gravitational, electroweak and strong interactions is still missing. A second difficulty arises in the need to introduce dark matter and dark energy to reproduce rotation curves of galaxies and the accelerating expansion of the Universe, in the so-called $\Lambda$CDM cosmological model of the Universe. In this model, dark matter and dark energy introduced in an \emph{ad hoc} manner represent $95\%$ of mass and energy in the Universe.

In this context, many experiments and many observations are used to test fundamental physical laws. One can refer for instance to \cite{Safronova2018}\cite{Will2014}\cite{Uzan2011}\cite{Reynaud2009}\cite{Mattingly2005} for an overview. Here, we will highlight tests using highly accurate atomic frequency standards. Unlike tests based on observations over geological and cosmological timescales, tests using atomic frequency standards are present day laboratory experiments which do not require a particular cosmological model for their interpretation. Also, they can be repeated for verification purposes. For these reasons, they are good tools for revealing physics beyond general relativity and the standard model and for providing experimental constraints to guide the development of unified theories (see for instance \cite{Marciano1984}\cite{Damour1994}\cite{Damour2012}).

Within its uncertainty, an atomic frequency standard give access to the unperturbed atomic frequency of the chosen transition. Comparisons of atomic frequency standards based on different atomic transitions determine atomic frequency ratios. These frequency ratios are dimensionless quantities given by nature, independently of any system of units. A variation of atomic frequency ratios would be violation of the framework defined by general relativity and the standard model of particle physics and of its founding principles. This can be tested with repeated measurements of atomic frequency ratios. Time series of such measurements are used to search for signals indicative of variations induced by putative phenomena. Linear variations with time could be a present day effect of evolution of the Universe over cosmological timescale. Variations synchronous with the Earth's motion around the Sun could be an effect of an extraneous coupling to the gravitational field of the Sun or to additional field(s) finding its source in the Sun. Sinusoidal variations could be due to an extraneous coupling to certain candidate dark matter field condensed on our galaxy. For example, linear variations of the fine-structure constant $\alpha$, of the electron-to-proton mass ratio $\mu=m_e/m_p$ and of the quark mass $m_q/\Lambda_{\mathrm{QCD}}$ are tested with uncertainties of $2.3\times 10^{-17}$~yr$^{-1}$, $7.5\times 10^{-17}$~yr$^{-1}$ and $1.8\times 10^{-15}$~yr$^{-1}$. Variations with respect to gravity (null redshift test) are tested with uncertainties of $1.0\times 10^{-7}$, $8.8\times 10^{-7}$ and $2.3\times 10^{-6}$ (see, for instance, \cite{Guena2012a}\cite{Safronova2018} and references therein). Searches for dark matter with frequency standards are reported in \cite{VanTilburg2015a}\cite{Hees2016}\cite{Wcislo2016a}. All these tests improve regularly with the steady progress of atomic frequency standards.

Time and frequency metrology provide many other tests of local position invariance and local Lorentz invariance. New possibilities are frequently emerging, like for instance, use of fiber links \cite{Delva2017a} for special relativity tests, of GALILEO satellites for gravitational redshift test \cite{Delva2018}. The ACES space mission \cite{Meynadier2018}\cite{Laurent2015}\cite{Cacciapuoti2009}\cite{Cacciapuoti2007} will provide improved tests and new opportunities. It is worth noting here that the third component of Einstein's equivalence principle, the universality of free-fall, was recently tested with improved uncertainty with the MICROSCOPE space mission (see \cite{Touboul2017} and references therein).

\section{Toward a redefinition of the SI second}\label{sec_redefinition}

Optical frequency standards now surpass $^{133}$Cs fountain primary standards by more than 2 orders of magnitude, both in stability and uncertainty. This naturally creates a strong incentive to redefine the SI second based on optical transition(s). The advent of this possibility was anticipated shortly after the demonstration of the first optical frequency comb. In 2001, Consultative Committee for Time and Frequency (CCTF) recommended that a list of secondary representations of the second (SRS) be established \cite{CCTF2001}, as one of the key processes to engage into a possible redefinition of the second. In 2004, International Committee for Weights and Measures (CIPM) adopted the first SRS, the $^{87}$Rb hyperfine frequency based on developments and measurements made at LNE-SYRTE \cite{Marion2003}\cite{Bize2004}. To date, the list of SRS comprises 9 transitions, 8 of which are optical transitions \cite{Riehle2018}. SRS are atomic transitions used to realize frequency standards with excellent uncertainties and which are measured in the SI system with accuracies close to the limit of $^{133}$Cs fountain standards. The recommended fractional frequency uncertainties of SRS range from $4\times 10^{-16}$ to $1.9\times 10^{-15}$ in the last version of the list of recommended standard frequencies of the CIPM \cite{BIPM_LOR}\cite{Riehle2018}. The recommended values of SRS come from the work of the joint CCL-CCTF Frequency Standard working group of the CIPM. Values are determined based on the least-square adjustments to high accuracy measurements found in peer-reviewed publications \cite{Margolis2015}\cite{Robertsson2016}\cite{Riehle2018}. This work enables checking the consistency of published measurements and the status of atomic frequency standards with the lowest uncertainties.

In view of a possible redefinition of the second, a key aspect of the process is to link the optical frequency domain to the present definition. In the current status of the list of recommended standard frequencies, this is achieved by some 55 highly accurate absolute frequency measurements of optical transitions. The 698~nm $^{87}$Sr $^1$S$_0$-$^3$P$_0$ transition used in optical lattice standards gathers by far the largest number of measurements and the ones with the highest accuracy from LNE-SYRTE \cite{LeTargat2013a}\cite{Lodewyck2016} and PTB \cite{Grebing2016}. These measurements are limited by the accuracy of best realizations of the second based on the $^{133}$Cs hyperfine transition. Measurements used to establish the list of recommended values also include a still limited number of optical-to-optical frequency ratios (8 measurements of 5 ratios). Such measurements are not limited by $^{133}$Cs fountain. The best measurements so far have a fractional frequency uncertainties near 5 parts in $10^{17}$ \cite{Rosenband2008}\cite{Nemitz2016a}. Among optical-to-optical frequency ratios, only the $^{199}$Hg/$^{87}$Sr is measured independently in at least two laboratories (RIKEN~\cite{Yamanaka2015a} and LNE-SYRTE~\cite{Tyumenev2016}) with uncertainties below the current realization of the SI second. Optical-to-optical ratios measured independently provide means to check the status of optical frequency metrology beyond the limit imposed by $^{133}$Cs fountains. On-going work shall increase the number of such tests and push their uncertainties near 1 part in $10^{18}$.

A redefinition of the SI second must not have negative consequences on the elaboration of TAI. This implies that sustainable capability of calibrating TAI scale interval against optical transition(s) must be proven before the redefinition. Most of the current TAI architecture and infrastructure operates in the microwave domain and it will remain so for many more years. Combs enable dividing the frequency of an optical reference to the microwave and RF domains with a limited degradation of its accuracy. An optical frequency standard can thereby readily be inserted in lieu of $^{133}$Cs fountains in the TAI calibration processes of metrology institutes, provided that its reference transition is adopted as a secondary representation of the second \cite{Guena2014}. The minimum requirement is to being able to calibrate the mean frequency of a local oscillator linked to TAI (typically, a hydrogen maser) over 5 days of a conventional grid. This was pioneered by LNE-SYRTE using the $^{87}$Sr $^1$S$_0$-$^3$P$_0$ transition. A first series of TAI calibrations with this SRS is reported in the BIPM \emph{Circular~T}350 \cite{CircularT}. These first calibrations were not used to steering the frequency of TAI. At the beginning of 2019, \emph{Circular~T}372 was published with the first TAI calibrations by optical standards used for steering, still with $^{87}$Sr. Reliable operation and use of $^{87}$Sr optical lattice standards to this end is described in \cite{Lodewyck2016}\cite{Hachisu2018}. Along the same line, application of a $^{87}$Sr optical lattice standard for realizing a representative sample local timescale is reported in \cite{Grebing2016}. Remote comparison of $^{87}$Sr optical lattice standards over intercontinental baseline via satellite-based methods and for almost 1~day is reported in \cite{Hachisu2014a}. In both cases, the architecture of the timescale and links is in the microwave domain. It remains a topic of investigation to define novel architectures and infrastructures that can take full benefit of the 2 orders of magnitude improved characteristics of optical frequency standards \cite{Riehle2017}. Such architectures and infrastructures may include optical fibers links, optical local oscillators, optical clockwork like the one demonstrated in \cite{Herman2018}, and novel methods.

Altogether, CCTF, in its strategy document, keeps a list of requirements for a redefinition of the second based on optical transition(s) to become possible. The aim is to make a choice that will last long, to ensure continuity, to guaranty gapless dissemination in particular via the elaboration of TAI and to validate the uncertainty of optical frequency standards \cite{Riehle2018}. CCTF and the joint CCL-CCTF Frequency Standard working group monitor progress of the field in this direction.

\section{Chronometric geodesy}\label{sec_geodesy}

Remote comparisons between two identical frequency standards show a frequency ratio $\nu_2/\nu_1=[1-(U_2-U_1)/c^2]$ where $U_1$ and $U_2$ are gravitational potentials at the location of the two standards. This gravitational redshift already mentioned in section~\ref{sec_TAI} amounts to $\sim 10^{-18}$ per centimeter of elevation in the vicinity of the Earth's surface. Frequency standards can be viewed as sensors to determine gravitational potential differences for the purpose of geodesy and Earth's science. This method called chronometric geodesy was proposed several decades ago (see for example \cite{Delva2019} and references therein). However, it is only now that progress in optical frequency standards and in optical frequency metrology makes this method potentially relevant.

A few years ago, chronometric geodesy gained significant interest from both time and frequency metrology and geodesy communities. Several aspects need being covered. One is to develop methodologies to use data from frequency standards in conjunction with other data already used in gravity field modelling. In this context, optical frequency standards can be considered as sensors of the potential with a point-like spatial response function, since the spread of the atomic sample is less 1~mm. Instead, space-based gravity measurements (GRACE, GOCE missions) gives 1$^{\mathrm{st}}$ or higher derivatives of the potential with a spatial response function with characteristic size of 100~km. Gravimeters and gradiometers on ground give the 1$^{\mathrm{st}}$ and 2$^{\mathrm{nd}}$ derivatives of the potential respectively with a point-like response ($0.1$~m$-1$~m), and so on. Geodetic methods to link the potential at the location of a frequency standard to larger and global scale are described in \cite{Lisdat2016}\cite{Denker2017}\cite{Denker2013}. These methods were applied to several places in Europe and provided improvements over previous similar determination \cite{Weiss2017}\cite{Calonico2007}\cite{Pavlis2003}. Specific gravimetry and levelling measurements in the vicinity (up to few 10~km) of frequency standard location are required to connect to the regional and global scale to the $10^{-18}$ or equivalently to the $1$~cm level. Time variations must be taken into account since tides can produce local gravity potential changes of up to $10^{-16}$. This is described in \cite{Voigt2016}.

Another aspect of investigations of chronometric geodesy is to understand and define cases where data from optical frequency standards can bring the most to applications in geodesy and geophysics. Reference \cite{Lion2017} reports one of the first quantitative studies along this line. Representative cases of hilly areas were studied to determine how much adding data from optical frequency standards can improve high spatial resolution gravity modelling. A fully synthetic simulation framework was developed that will be further refined and applied to other cases of potential interest, like for instance, coastal areas. Interest can be two fold: improvement of reference systems and studies of geophysical phenomena. Some possibilities are discussed in \cite{Mehlstaubler2018}\cite{Muller2018}\cite{Bondarescu2015a}\cite{Bondarescu2012}. Within the last few years, working groups within the International Association of Geodesy (IAG) were initiated to consider potentialities of chronometric geodesy.

A third aspect is the further development of instrumental capabilities. The long term goal would be to have ruggedized field-compatible optical frequency standards with uncertainties $\leq 10^{-18}$ and means of comparing them from any place of interest on Earth. First proof-of-concept experiments used coherent optical fiber links between laboratories with already existing state-of-the-art optical frequency metrology programs \cite{Lisdat2016}\cite{Takano2016}. Over the years, transportable optical frequency standards were developed and currently achieve uncertainties in the $10^{-17}$ range \cite{Bongs2015a}\cite{Koller2017}\cite{Origlia2018}, also with the aim to increase technology readiness for space. Recently, one of these devices was used for another proof-of-concept experiment \cite{Grotti2018} again in conjunction with coherent optical fiber link. Solutions for remote comparisons that reach $10^{-18}$ firstly without optical fiber link and secondly over intercontinental distances are still missing. For the former, free-space optical links studied in \cite{Cossel2017}\cite{Deschenes2016}\cite{Giorgetta2013a}\cite{Djerroud2010} could be extended to a few 100~km distances with the help of an airborne platform.

The further advancement of all the above studies shall lead to first pilot chronometric geodesy programs designed for geodesy, geophysics and Earth's science applications.

%

\section{Conclusions}

We described the status of the realization of the SI unit of time according to its current definition. We reported on the tremendous progress of optical frequency metrology over the last 20~years. We mentioned how the CIPM monitors the situation of the field in view of a possible redefinition. To date, work remains to be done to meet all milestones of the CCTF strategy document and to prepare for long-term commitments of the post redefinition era. When this will be achieved, progress of frequency and time metrology and its applications will not need being limited by $^{133}$Cs standards any more. Secondary representations of the second and highly accurate redundant measurements of a sufficient number of optical-to-optical frequency ratios will permit a smooth passage through the redefinition. On the basis of these considerations and taking into account the time for due process at the CIPM and at the CGPM, a redefinition seems possible before 2030. The new definition must last long. Paradoxically, the persistent vitality of research on optical frequency standards could be a reason to delay the redefinition, if new breakthroughs suggest even better choices than those being currently consolidated.
One possible choice for the redefinition would be to mimic the present definition and select one single optical transition. In this case, secondary representations will be useful, even needed in practice. The possibility, the shape and the relevance of a system using several transitions on an even basis remain to be explored.

The choice of one particular transition of one particular atom (or a set of them) remains a kind of artefact.
Atoms are already complex assemblage of elementary particles and transition frequencies are by far not calculable at the level of uncertainty to which they can be realized by atomic frequency standards. The article of C.~Bord\'e in the present issue of Comptes Rendus presented the deeper foundations of the international system of units, its links with geometry of space-time and with the system of units introduced by Planck based on five fundamental constants. The system adopted by the 26$^{\mathrm{th}}$ CGPM is based on fixing $\Delta\nu_{\mathrm{Cs}}$, $c$, $h$, $e$, $k_{\mathrm{B}}$.
Moving to a system of units founded solely on most fundamental aspects of physical laws would require to abandon atomic transition(s) for the definition of the unit of time. Referring to Planck's system, that would mean fixing the gravitational constant $G$ to define Planck's time $\sqrt{\hbar G/c^5}$. Another somewhat intermediate possibility could be to fix the mass of an elementary particle, for instance, the electron $m_e$ which defines the unit of time as $h/m_e c^2$. Given the utmost importance of practical aspects of the actual implementation of the SI system, it does not seem presently possible to opt for one of these fundamental definitions of time because we do not know how to realize them with the same exquisite accuracy of atomic frequency standards. Research shall investigate how to measure $G$ or $m_e$ with radically improved accuracy in the SI system of units attached to atomic transitions.

\section{Acknowledgements}

The author is grateful to C. Salomon for his patience, his careful reading of the manuscript and his suggestions to improve it.

\section{References}


\bibliography{D:/Bibliographie/BiblioLatex/bibliography}

\begin{thebibliography}{100}
\expandafter\ifx\csname url\endcsname\relax
  \def\url#1{\texttt{#1}}\fi
\expandafter\ifx\csname urlprefix\endcsname\relax\def\urlprefix{URL }\fi
\expandafter\ifx\csname href\endcsname\relax
  \def\href#1#2{#2} \def\path#1{#1}\fi

\bibitem{CGPM1967}
\href{https://www.bipm.org/en/worldwide-metrology/cgpm/resolutions.html}{Comptes
  rendus de la 13i\`eme {CGPM} (1967/68)} (1969) 103.
\newline\urlprefix\url{https://www.bipm.org/en/worldwide-metrology/cgpm/resolutions.html}

\bibitem{Lyons1952}
H.~Lyons, Spectral {{Lines}} as {{Frequency Standards}}, Annals of the New York
  Academy of Sciences 55~(5) (1952) 831--871.
\newblock \href {http://dx.doi.org/10.1111/j.1749-6632.1952.tb26600.x}
  {\path{doi:10.1111/j.1749-6632.1952.tb26600.x}}.

\bibitem{Essen1955}
J.~V.~L. Essen, L.and~Parry, \href{http://dx.doi.org/10.1038/176280a0}{An
  atomic standard of frequency and time interval: A caesium resonator}, Nature
  176~(4476) (1955) 280--282.
\newline\urlprefix\url{http://dx.doi.org/10.1038/176280a0}

\bibitem{Markowitz1958}
W.~Markowitz, R.~G. Hall, L.~Essen, J.~V.~L. Parry,
  \href{http://link.aps.org/doi/10.1103/PhysRevLett.1.105}{Frequency of cesium
  in terms of ephemeris time}, Phys. Rev. Lett. 1 (1958) 105--107.
\newblock \href {http://dx.doi.org/10.1103/PhysRevLett.1.105}
  {\path{doi:10.1103/PhysRevLett.1.105}}.
\newline\urlprefix\url{http://link.aps.org/doi/10.1103/PhysRevLett.1.105}

\bibitem{Quinn2005}
T.~Quinn, Fifty years of atomic time-keeping: 1955 to 2005, Metrologia 42~(3).
\newblock \href {http://dx.doi.org/10.1088/0026-1394/42/3/E01}
  {\path{doi:10.1088/0026-1394/42/3/E01}}.

\bibitem{Salomon2015}
C.~Salomon, The measurement of time / {{La}} mesure du temps: {{Foreword}},
  Comptes Rendus Physique 16~(5) (2015) 459--460.
\newblock \href {http://dx.doi.org/10.1016/j.crhy.2015.05.006}
  {\path{doi:10.1016/j.crhy.2015.05.006}}.

\bibitem{CGPM1983}
Comptes rendus de la 17i\`eme {CGPM} (1983) 10.

\bibitem{CGPM2018}
Comptes rendus de la 26i\`eme {CGPM}.

\bibitem{Uzan2011}
J.-P. Uzan, \href{http://www.livingreviews.org/lrr-2011-2}{Varying constants,
  gravitation and cosmology}, Living Reviews in Relativity 14~(2).
\newline\urlprefix\url{http://www.livingreviews.org/lrr-2011-2}

\bibitem{Will2014}
C.~M. Will, \href{https://doi.org/10.12942/lrr-2014-4}{The confrontation
  between general relativity and experiment}, Living Reviews in Relativity
  17~(1) (2014) 4.
\newblock \href {http://dx.doi.org/10.12942/lrr-2014-4}
  {\path{doi:10.12942/lrr-2014-4}}.
\newline\urlprefix\url{https://doi.org/10.12942/lrr-2014-4}

\bibitem{Safronova2018}
M.~S. Safronova, D.~Budker, D.~DeMille, D.~F.~J. Kimball, A.~Derevianko, C.~W.
  Clark, Search for new physics with atoms and molecules, Rev. Mod. Phys.
  90~(2) (2018) 025008.
\newblock \href {http://dx.doi.org/10.1103/RevModPhys.90.025008}
  {\path{doi:10.1103/RevModPhys.90.025008}}.

\bibitem{VanTilburg2015a}
K.~Van~Tilburg, N.~Leefer, L.~Bougas, D.~Budker,
  \href{http://link.aps.org/doi/10.1103/PhysRevLett.115.011802}{Search for
  ultralight scalar dark matter with atomic spectroscopy}, Phys. Rev. Lett. 115
  (2015) 011802.
\newblock \href {http://dx.doi.org/10.1103/PhysRevLett.115.011802}
  {\path{doi:10.1103/PhysRevLett.115.011802}}.
\newline\urlprefix\url{http://link.aps.org/doi/10.1103/PhysRevLett.115.011802}

\bibitem{Hees2016}
A.~Hees, J.~Gu\'ena, M.~Abgrall, S.~Bize, P.~Wolf,
  \href{http://link.aps.org/doi/10.1103/PhysRevLett.117.061301}{Searching for
  an oscillating massive scalar field as a dark matter candidate using atomic
  hyperfine frequency comparisons}, Phys. Rev. Lett. 117 (2016) 061301.
\newblock \href {http://dx.doi.org/10.1103/PhysRevLett.117.061301}
  {\path{doi:10.1103/PhysRevLett.117.061301}}.
\newline\urlprefix\url{http://link.aps.org/doi/10.1103/PhysRevLett.117.061301}

\bibitem{Ramsey1990}
N.~Ramsey, Experiments with separated oscillatory fields and hydrogen masers,
  Rev. Mod. Phys. 62 (1990) 541.
\newblock \href {http://dx.doi.org/https://doi.org/10.1103/RevModPhys.62.541}
  {\path{doi:https://doi.org/10.1103/RevModPhys.62.541}}.

\bibitem{Vanier1989}
J.~Vanier, C.~Audoin, The Quantum Physics of Atomic Frequency Standards, Adam
  Hilger, 1989.

\bibitem{CohenTannoudji1998}
C.~N. Cohen-Tannoudji,
  \href{http://link.aps.org/doi/10.1103/RevModPhys.70.707}{Nobel lecture:
  Manipulating atoms with photons}, Rev. Mod. Phys. 70 (1998) 707--719.
\newblock \href {http://dx.doi.org/10.1103/RevModPhys.70.707}
  {\path{doi:10.1103/RevModPhys.70.707}}.
\newline\urlprefix\url{http://link.aps.org/doi/10.1103/RevModPhys.70.707}

\bibitem{Phillips1998}
W.~D. Phillips, Nobel lecture: Laser cooling and trapping of neutral atoms,
  Rev. Mod. Phys. 70~(3) (1998) 721--741.
\newblock \href {http://dx.doi.org/10.1103/RevModPhys.70.721}
  {\path{doi:10.1103/RevModPhys.70.721}}.

\bibitem{Chu1998}
S.~Chu, \href{http://link.aps.org/doi/10.1103/RevModPhys.70.685}{Nobel lecture:
  The manipulation of neutral particles}, Rev. Mod. Phys. 70 (1998) 685--706.
\newblock \href {http://dx.doi.org/10.1103/RevModPhys.70.685}
  {\path{doi:10.1103/RevModPhys.70.685}}.
\newline\urlprefix\url{http://link.aps.org/doi/10.1103/RevModPhys.70.685}

\bibitem{Kasevich1989}
M.~Kasevich, E.~Riis, S.~Chu, R.~de~Voe, {RF} spectroscopy in an atomic
  fountain, Phys. Rev. Lett. 63 (1989) 612.

\bibitem{Clairon1991}
A.~Clairon, C.~Salomon, S.~Guellati, W.~Phillips, {Ramsey resonance in a
  Zacharias fountain}, Europhys. Lett. 16 (1991) 165.

\bibitem{Clairon1995}
A.~Clairon, P.~Laurent, G.~Santarelli, S.~Ghezali, S.~Lea, M.~Bahoura, A cesium
  fountain frequency standard: recent results, IEEE Trans. on Inst. and Meas.
  44~(2) (1995) 128.

\bibitem{Santarelli1999}
G.~Santarelli, P.~Laurent, P.~Lemonde, A.~Clairon, A.~G. Mann, S.~Chang, A.~N.
  Luiten, C.~Salomon, Quantum projection noise in an atomic fountain: A high
  stability cesium frequency standard, Phys. Rev. Lett. 82~(23) (1999) 4619.

\bibitem{Guena2012}
J.~Gu{\'e}na, M.~Abgrall, D.~Rovera, P.~Laurent, B.~Chupin, M.~Lours,
  G.~Santarelli, P.~Rosenbusch, M.~Tobar, R.~Li, K.~Gibble, A.~Clairon,
  S.~Bize, \href{https://doi.org/10.1109/TUFFC.2012.2208}{Progress in atomic
  fountains at {LNE-SYRTE}}, Ultrasonics, Ferroelectrics and Frequency Control,
  IEEE Transactions on 59~(3) (2012) 391--410.
\newblock \href {http://dx.doi.org/10.1109/TUFFC.2012.2208}
  {\path{doi:10.1109/TUFFC.2012.2208}}.
\newline\urlprefix\url{https://doi.org/10.1109/TUFFC.2012.2208}

\bibitem{Guena2011}
J.~Gu{\'e}na, R.~Li, K.~Gibble, S.~Bize, A.~Clairon,
  \href{https://doi.org/10.1103/PhysRevLett.106.130801}{Evaluation of {D}oppler
  shifts to improve the accuracy of primary atomic fountain clocks}, Phys. Rev.
  Lett. 106~(13) (2011) 130801.
\newblock \href {http://dx.doi.org/10.1103/PhysRevLett.106.130801}
  {\path{doi:10.1103/PhysRevLett.106.130801}}.
\newline\urlprefix\url{https://doi.org/10.1103/PhysRevLett.106.130801}

\bibitem{Li2010b}
R.~Li, K.~Gibble,
  \href{http://stacks.iop.org/0026-1394/47/i=5/a=004}{Evaluating and minimizing
  distributed cavity phase errors in atomic clocks}, Metrologia 47~(5) (2010)
  534.
\newline\urlprefix\url{http://stacks.iop.org/0026-1394/47/i=5/a=004}

\bibitem{Pereira2002}
F.~{Pereira Dos Santos}, H.~Marion, S.~Bize, Y.~Sortais, A.~Clairon,
  C.~Salomon, Controlling the cold collision shift in high precision atomic
  interferometry, Phys. Rev. Lett. 89 (2002) 233004.

\bibitem{Papoular2012a}
D.~J. Papoular, S.~Bize, A.~Clairon, H.~Marion, S.~J. J. M.~F. Kokkelmans,
  G.~V. Shlyapnikov,
  \href{http://link.aps.org/doi/10.1103/PhysRevA.86.040701}{Feshbach resonances
  in cesium at ultralow static magnetic fields}, Phys. Rev. A 86 (2012) 040701.
\newblock \href {http://dx.doi.org/10.1103/PhysRevA.86.040701}
  {\path{doi:10.1103/PhysRevA.86.040701}}.
\newline\urlprefix\url{http://link.aps.org/doi/10.1103/PhysRevA.86.040701}

\bibitem{Szymaniec2007}
K.~Szymaniec, W.~Cha\l{}upczak, E.~Tiesinga, C.~J. Williams, S.~Weyers,
  R.~Wynands, Cancellation of the collisional frequency shift in caesium
  fountain clocks, Phys. Rev. Lett. 98~(15) (2007) 153002.
\newblock \href {http://dx.doi.org/10.1103/PhysRevLett.98.153002}
  {\path{doi:10.1103/PhysRevLett.98.153002}}.

\bibitem{Gibble2013}
K.~Gibble,
  \href{http://link.aps.org/doi/10.1103/PhysRevLett.110.180802}{Scattering of
  cold-atom coherences by hot atoms: Frequency shifts from background-gas
  collisions}, Phys. Rev. Lett. 110 (2013) 180802.
\newblock \href {http://dx.doi.org/10.1103/PhysRevLett.110.180802}
  {\path{doi:10.1103/PhysRevLett.110.180802}}.
\newline\urlprefix\url{http://link.aps.org/doi/10.1103/PhysRevLett.110.180802}

\bibitem{Szymaniec2014}
K.~Szymaniec, S.~Lea, K.~Liu, An evaluation of the frequency shift caused by
  collisions with background gas in the primary frequency standard {NPL-CsF2},
  Ultrasonics, Ferroelectrics and Frequency Control, IEEE Transactions on
  61~(1) (2014) 203--206.
\newblock \href {http://dx.doi.org/10.1109/TUFFC.2014.6689789}
  {\path{doi:10.1109/TUFFC.2014.6689789}}.

\bibitem{Gibble2006}
K.~Gibble, \href{http://link.aps.org/abstract/PRL/v97/e073002}{Difference
  between a photon's momentum and an atom's recoil}, Phys. Rev. Lett. 97~(7)
  (2006) 073002.
\newblock \href {http://dx.doi.org/10.1103/PhysRevLett.97.073002}
  {\path{doi:10.1103/PhysRevLett.97.073002}}.
\newline\urlprefix\url{http://link.aps.org/abstract/PRL/v97/e073002}

\bibitem{Li2011}
R.~Li, K.~Gibble, K.~Szymaniec,
  \href{http://stacks.iop.org/0026-1394/48/i=5/a=007}{Improved accuracy of the
  {NPL}-{CsF2} primary frequency standard: evaluation of distributed cavity
  phase and microwave lensing frequency shifts}, Metrologia 48~(5) (2011) 283.
\newline\urlprefix\url{http://stacks.iop.org/0026-1394/48/i=5/a=007}

\bibitem{Gibble2014}
K.~Gibble, \href{http://link.aps.org/doi/10.1103/PhysRevA.90.015601}{Ramsey
  spectroscopy, matter-wave interferometry, and the microwave-lensing frequency
  shift}, Phys. Rev. A 90 (2014) 015601.
\newblock \href {http://dx.doi.org/10.1103/PhysRevA.90.015601}
  {\path{doi:10.1103/PhysRevA.90.015601}}.
\newline\urlprefix\url{http://link.aps.org/doi/10.1103/PhysRevA.90.015601}

\bibitem{Itano1982}
W.~Itano, L.~Lewis, D.~Wineland, Shift of $^{2}${S}$_{1/2}$ hyperfine
  splittings due to blackbody radiation, Phys. Rev. A. 25 (1982) 1233.

\bibitem{Jefferts2014}
R.~Jefferts, S.\, P.~Heavner, T.\, E.~Parker, T.\, H.~Shirley, J.\, A.~Donley,
  E.\, N.~Ashby, F.~Levi, D.~Calonico, G.~A. Costanzo,
  \href{http://link.aps.org/doi/10.1103/PhysRevLett.112.050801}{High-accuracy
  measurement of the blackbody radiation frequency shift of the ground-state
  hyperfine transition in $^{133}${Cs}}, Phys. Rev. Lett. 112 (2014) 050801.
\newblock \href {http://dx.doi.org/10.1103/PhysRevLett.112.050801}
  {\path{doi:10.1103/PhysRevLett.112.050801}}.
\newline\urlprefix\url{http://link.aps.org/doi/10.1103/PhysRevLett.112.050801}

\bibitem{Heavner2014}
T.~P. Heavner, E.~A. Donley, F.~Levi, G.~Costanzo, T.~E. Parker, J.~H. Shirley,
  N.~Ashby, S.~Barlow, S.~R. Jefferts,
  \href{http://stacks.iop.org/0026-1394/51/i=3/a=174}{First accuracy evaluation
  of {NIST-F2}}, Metrologia 51~(3) (2014) 174.
\newline\urlprefix\url{http://stacks.iop.org/0026-1394/51/i=3/a=174}

\bibitem{Levi2014}
F.~Levi, D.~Calonico, C.~E. Calosso, A.~Godone, S.~Micalizio, G.~A. Costanzo,
  \href{http://stacks.iop.org/0026-1394/51/i=3/a=270}{Accuracy evaluation of
  {ITCsF2}: a nitrogen cooled caesium fountain}, Metrologia 51~(3) (2014) 270.
\newline\urlprefix\url{http://stacks.iop.org/0026-1394/51/i=3/a=270}

\bibitem{Jallageas2018}
A.~Jallageas, L.~Devenoges, M.~Petersen, J.~Morel, L.~G. Bernier, D.~Schenker,
  P.~Thomann, T.~S\"udmeyer, First uncertainty evaluation of the {{FoCS}}-2
  primary frequency standard, Metrologia 55~(3) (2018) 366.
\newblock \href {http://dx.doi.org/10.1088/1681-7575/aab3fa}
  {\path{doi:10.1088/1681-7575/aab3fa}}.

\bibitem{Wynands2005}
R.~Wynands, S.~Weyers,
  \href{http://stacks.iop.org/0026-1394/42/i=3/a=S08}{Atomic fountain clocks},
  Metrologia 42~(3) (2005) S64.
\newline\urlprefix\url{http://stacks.iop.org/0026-1394/42/i=3/a=S08}

\bibitem{Riehle2016}
F.~Riehle, \href{http://stacks.iop.org/1742-6596/723/i=1/a=011001}{8th
  symposium on frequency standards and metrology 2015}, Journal of Physics:
  Conference Series 723~(1) (2016) 011001.
\newline\urlprefix\url{http://stacks.iop.org/1742-6596/723/i=1/a=011001}

\bibitem{Guena2010}
J.~Gu{\'e}na, P.~Rosenbusch, P.~Laurent, M.~Abgrall, D.~Rovera, G.~Santarelli,
  M.~Tobar, S.~Bize, A.~Clairon,
  \href{https://doi.org/10.1109/TUFFC.2010.1461}{Demonstration of a dual alkali
  {R}b/{C}s fountain clock}, Ultrasonics, Ferroelectrics and Frequency Control,
  IEEE Transactions on 57~(3) (2010) 647.
\newblock \href {http://dx.doi.org/10.1109/TUFFC.2010.1461}
  {\path{doi:10.1109/TUFFC.2010.1461}}.
\newline\urlprefix\url{https://doi.org/10.1109/TUFFC.2010.1461}

\bibitem{Guena2014}
J.~Gu\'ena, M.~Abgrall, A.~Clairon, S.~Bize,
  \href{http://stacks.iop.org/0026-1394/51/i=1/a=108}{Contributing to {TAI}
  with a secondary representation of the si second}, Metrologia 51~(1) (2014)
  108.
\newline\urlprefix\url{http://stacks.iop.org/0026-1394/51/i=1/a=108}

\bibitem{BIPM_LOR}
See the list of Secondary Representations of the SI second and of other
  recommended values of standard frequencies on the BIPM website.
\newblock \href{http://www.bipm.org/en/publications/mep.html}{[link]}.
\newline\urlprefix\url{http://www.bipm.org/en/publications/mep.html}

\bibitem{Bauch2006}
A.~Bauch, J.~Achkar, S.~Bize, D.~Calonico, R.~Dach, R.~Hlavac, L.~Lorini,
  T.~Parker, G.~Petit, D.~Piester, K.~Szymaniec, P.~Uhrich,
  \href{http://stacks.iop.org/0026-1394/43/109}{Comparison between frequency
  standards in europe and the {USA} at the $10^{-15}$ uncertainty level},
  Metrologia 43~(1) (2006) 109--120.
\newline\urlprefix\url{http://stacks.iop.org/0026-1394/43/109}

\bibitem{Guena2017}
J.~Guéna, S.~Weyers, M.~Abgrall, C.~Grebing, V.~Gerginov, P.~Rosenbusch,
  S.~Bize, B.~Lipphardt, H.~Denker, N.~Quintin, S.~M.~F. Raupach, D.~Nicolodi,
  F.~Stefani, N.~Chiodo, S.~Koke, A.~Kuhl, F.~Wiotte, F.~Meynadier,
  E.~Camisard, C.~Chardonnet, Y.~L. Coq, M.~Lours, G.~Santarelli, A.~Amy-Klein,
  R.~L. Targat, O.~Lopez, P.~E. Pottie, G.~Grosche,
  \href{http://stacks.iop.org/0026-1394/54/i=3/a=348}{First international
  comparison of fountain primary frequency standards via a long distance
  optical fiber link}, Metrologia 54~(3) (2017) 348.
\newline\urlprefix\url{http://stacks.iop.org/0026-1394/54/i=3/a=348}

\bibitem{Petit2015a}
G.~Petit, F.~Arias, G.~Panfilo,
  \href{http://www.sciencedirect.com/science/article/pii/S1631070515000535}{International
  atomic time: Status and future challenges}, Comptes Rendus Physique 16~(5)
  (2015) 480 -- 488, the measurement of time / La mesure du temps.
\newblock \href
  {http://dx.doi.org/http://dx.doi.org/10.1016/j.crhy.2015.03.002}
  {\path{doi:http://dx.doi.org/10.1016/j.crhy.2015.03.002}}.
\newline\urlprefix\url{http://www.sciencedirect.com/science/article/pii/S1631070515000535}

\bibitem{CircularT}
See \emph{Circular T} and fountain reports on the BIPM website.
\newblock
  \href{https://www.bipm.org/fr/bipm-services/timescales/time-ftp/Circular-T.html}{[link]}.
\newline\urlprefix\url{https://www.bipm.org/fr/bipm-services/timescales/time-ftp/Circular-T.html}

\bibitem{Parker2012}
T.~E. Parker, \href{http://link.aip.org/link/?RSI/83/021102/1}{Invited review
  article: The uncertainty in the realization and dissemination of the {SI}
  second from a systems point of view}, Review of Scientific Instruments 83~(2)
  (2012) 021102.
\newblock \href {http://dx.doi.org/10.1063/1.3682002}
  {\path{doi:10.1063/1.3682002}}.
\newline\urlprefix\url{http://link.aip.org/link/?RSI/83/021102/1}

\bibitem{Petit2013}
G.~Petit, G.~Panfilo, Comparison of frequency standards used for {TAI},
  Instrumentation and Measurement, IEEE Transactions on 62~(99) (2013) 1550.
\newblock \href {http://dx.doi.org/10.1109/TIM.2012.2228749}
  {\path{doi:10.1109/TIM.2012.2228749}}.

\bibitem{Riehle2018}
F.~Riehle, P.~Gill, F.~Arias, L.~Robertsson,
  \href{http://stacks.iop.org/0026-1394/55/i=2/a=188}{The {CIPM} list of
  recommended frequency standard values: guidelines and procedures}, Metrologia
  55~(2) (2018) 188.
\newline\urlprefix\url{http://stacks.iop.org/0026-1394/55/i=2/a=188}

\bibitem{Hachisu2017}
H.~Hachisu, G.~Petit, F.~Nakagawa, Y.~Hanado, T.~Ido,
  \href{http://www.opticsexpress.org/abstract.cfm?URI=oe-25-8-8511}{{SI}-traceable
  measurement of an optical frequency at the low $10^{-16}$ level without a
  local primary standard}, Opt. Express 25~(8) (2017) 8511--8523.
\newblock \href {http://dx.doi.org/10.1364/OE.25.008511}
  {\path{doi:10.1364/OE.25.008511}}.
\newline\urlprefix\url{http://www.opticsexpress.org/abstract.cfm?URI=oe-25-8-8511}

\bibitem{McGrew2018a}
W.~F. McGrew, X.~Zhang, H.~Leopardi, R.~J. Fasano, D.~Nicolodi, K.~Beloy,
  J.~Yao, J.~A. Sherman, S.~A. Sch\"affer, J.~Savory, R.~C. Brown,
  S.~R\"omisch, C.~W. Oates, T.~E. Parker, T.~M. Fortier, A.~D. Ludlow, Towards
  {{Adoption}} of an {{Optical Second}}: {{Verifying Optical Clocks}} at the
  {{SI Limit}}, arXiv:1811.05885 [physics]\href
  {http://arxiv.org/abs/1811.05885} {\path{arXiv:1811.05885}}.

\bibitem{Ashby2007}
N.~Ashby, T.~P. Heavner, S.~R. Jefferts, T.~E. Parker, A.~G. Radnaev, Y.~O.
  Dudin, \href{http://link.aps.org/abstract/PRL/v98/e070802}{Testing {L}ocal
  {P}osition {I}nvariance with four cesium-fountain primary frequency standards
  and four {NIST} hydrogen masers}, Phys. Rev. Lett. 98 (2007) 070802.
\newblock \href {http://dx.doi.org/10.1103/PhysRevLett.98.070802}
  {\path{doi:10.1103/PhysRevLett.98.070802}}.
\newline\urlprefix\url{http://link.aps.org/abstract/PRL/v98/e070802}

\bibitem{Ashby2018}
N.~Ashby, T.~E. Parker, B.~R. Patla, A null test of general relativity based on
  a long-term comparison of atomic transition frequencies, Nature Physics
  (2018) 1\href {http://dx.doi.org/10.1038/s41567-018-0156-2}
  {\path{doi:10.1038/s41567-018-0156-2}}.

\bibitem{Bauch2012}
A.~Bauch, S.~Weyers, D.~Piester, E.~Staliuniene, W.~Yang,
  \href{http://stacks.iop.org/0026-1394/49/i=3/a=180}{Generation of {UTC(PTB)}
  as a fountain-clock based time scale}, Metrologia 49~(3) (2012) 180.
\newline\urlprefix\url{http://stacks.iop.org/0026-1394/49/i=3/a=180}

\bibitem{Peil2014}
S.~Peil, J.~L. Hanssen, T.~B. Swanson, J.~Taylor, C.~R. Ekstrom,
  \href{http://stacks.iop.org/0026-1394/51/i=3/a=263}{Evaluation of long term
  performance of continuously running atomic fountains}, Metrologia 51~(3)
  (2014) 263.
\newline\urlprefix\url{http://stacks.iop.org/0026-1394/51/i=3/a=263}

\bibitem{Rovera2016}
G.~D. Rovera, S.~Bize, B.~Chupin, J.~Gu\'ena, P.~Laurent, P.~Rosenbusch,
  P.~Uhrich, M.~Abgrall,
  \href{http://stacks.iop.org/0026-1394/53/i=3/a=S81}{{UTC(OP)} based on
  {LNE-SYRTE} atomic fountain primary frequency standards}, Metrologia 53~(3)
  (2016) S81.
\newline\urlprefix\url{http://stacks.iop.org/0026-1394/53/i=3/a=S81}

\bibitem{Petit2014}
G.~Petit, F.~Arias, A.~Harmegnies, G.~Panfilo, L.~Tisserand,
  \href{http://stacks.iop.org/0026-1394/51/i=1/a=33}{{UTCr}: a rapid
  realization of {UTC}}, Metrologia 51~(1) (2014) 33.
\newline\urlprefix\url{http://stacks.iop.org/0026-1394/51/i=1/a=33}

\bibitem{Hansch2006}
T.~W. Hansch, \href{http://link.aps.org/abstract/RMP/v78/p1297}{Nobel lecture:
  Passion for precision}, Reviews of Modern Physics 78~(4) (2006) 1297.
\newblock \href {http://dx.doi.org/10.1103/RevModPhys.78.1297}
  {\path{doi:10.1103/RevModPhys.78.1297}}.
\newline\urlprefix\url{http://link.aps.org/abstract/RMP/v78/p1297}

\bibitem{Hall2006}
J.~L. Hall, \href{http://link.aps.org/abstract/RMP/v78/p1279}{Nobel lecture:
  Defining and measuring optical frequencies}, Reviews of Modern Physics 78~(4)
  (2006) 1279.
\newblock \href {http://dx.doi.org/10.1103/RevModPhys.78.1279}
  {\path{doi:10.1103/RevModPhys.78.1279}}.
\newline\urlprefix\url{http://link.aps.org/abstract/RMP/v78/p1279}

\bibitem{McGrew2018b}
W.~F. McGrew, X.~Zhang, R.~J. Fasano, S.~A. Sch\"affer, K.~Beloy, D.~Nicolodi,
  R.~C. Brown, N.~Hinkley, G.~Milani, M.~Schioppo, T.~H. Yoon, A.~D. Ludlow,
  Atomic clock performance enabling geodesy below the centimetre level, Nature
  (2018) 1\href {http://dx.doi.org/10.1038/s41586-018-0738-2}
  {\path{doi:10.1038/s41586-018-0738-2}}.

\bibitem{Takano2016}
T.~Takano, M.~Takamoto, I.~Ushijima, N.~Ohmae, T.~Akatsuka, A.~Yamaguchi,
  Y.~Kuroishi, H.~Munekane, B.~Miyahara, H.~Katori,
  \href{http://dx.doi.org/10.1038/nphoton.2016.159}{Geopotential measurements
  with synchronously linked optical lattice clocks}, Nat Photon advance online
  publication.
\newline\urlprefix\url{http://dx.doi.org/10.1038/nphoton.2016.159}

\bibitem{Huntemann2016}
N.~Huntemann, C.~Sanner, B.~Lipphardt, C.~Tamm, E.~Peik,
  \href{http://link.aps.org/doi/10.1103/PhysRevLett.116.063001}{Single-ion
  atomic clock with $3\ifmmode\times\else\texttimes\fi{}{10}^{-18}$ systematic
  uncertainty}, Phys. Rev. Lett. 116 (2016) 063001.
\newblock \href {http://dx.doi.org/10.1103/PhysRevLett.116.063001}
  {\path{doi:10.1103/PhysRevLett.116.063001}}.
\newline\urlprefix\url{http://link.aps.org/doi/10.1103/PhysRevLett.116.063001}

\bibitem{Nicholson2015}
T.~L. Nicholson, S.~L. Campbell, R.~B. Hutson, G.~E. Marti, B.~J. Bloom, R.~L.
  McNally, W.~Zhang, M.~D. Barrett, M.~S. Safronova, G.~F. Strouse, W.~L. Tew,
  J.~Ye, \href{https://doi.org/10.1038/ncomms7896}{Systematic evaluation of an
  atomic clock at $2\times 10^{-18}$ total uncertainty}, Nature Communications
  6 (2015) 6896.
\newline\urlprefix\url{https://doi.org/10.1038/ncomms7896}

\bibitem{Ushijima2015}
I.~Ushijima, M.~Takamoto, M.~Das, T.~Ohkubo, H.~Katori,
  \href{http://dx.doi.org/10.1038/nphoton.2015.5}{Cryogenic optical lattice
  clocks}, Nat Photon 9 (2015) 185.
\newline\urlprefix\url{http://dx.doi.org/10.1038/nphoton.2015.5}

\bibitem{Bloom2014}
B.~J. Bloom, T.~L. Nicholson, J.~R. Williams, S.~L. Campbell, M.~Bishof,
  X.~Zhang, W.~Zhang, S.~L. Bromley, J.~Ye,
  \href{http://dx.doi.org/10.1038/nature12941}{An optical lattice clock with
  accuracy and stability at the $10^{-18}$ level}, Nature 506 (2014) 71.
\newline\urlprefix\url{http://dx.doi.org/10.1038/nature12941}

\bibitem{Beloy2014a}
K.~Beloy, N.~Hinkley, N.~B. Phillips, J.~A. Sherman, M.~Schioppo, J.~Lehman,
  A.~Feldman, L.~M. Hanssen, C.~W. Oates, A.~D. Ludlow,
  \href{http://link.aps.org/doi/10.1103/PhysRevLett.113.260801}{Atomic clock
  with $1\ifmmode\times\else\texttimes\fi{}{10}^{-18}$ room-temperature
  blackbody stark uncertainty}, Phys. Rev. Lett. 113 (2014) 260801.
\newblock \href {http://dx.doi.org/10.1103/PhysRevLett.113.260801}
  {\path{doi:10.1103/PhysRevLett.113.260801}}.
\newline\urlprefix\url{http://link.aps.org/doi/10.1103/PhysRevLett.113.260801}

\bibitem{Dube2014}
P.~Dub\'e, A.~A. Madej, M.~Tibbo, J.~E. Bernard,
  \href{http://link.aps.org/doi/10.1103/PhysRevLett.112.173002}{High-accuracy
  measurement of the differential scalar polarizability of a $^{88}$sr$^+$
  clock using the time-dilation effect}, Phys. Rev. Lett. 112 (2014) 173002.
\newblock \href {http://dx.doi.org/10.1103/PhysRevLett.112.173002}
  {\path{doi:10.1103/PhysRevLett.112.173002}}.
\newline\urlprefix\url{http://link.aps.org/doi/10.1103/PhysRevLett.112.173002}

\bibitem{Middelmann2012a}
T.~Middelmann, S.~Falke, C.~Lisdat, U.~Sterr,
  \href{http://link.aps.org/doi/10.1103/PhysRevLett.109.263004}{High accuracy
  correction of blackbody radiation shift in an optical lattice clock}, Phys.
  Rev. Lett. 109 (2012) 263004.
\newblock \href {http://dx.doi.org/10.1103/PhysRevLett.109.263004}
  {\path{doi:10.1103/PhysRevLett.109.263004}}.
\newline\urlprefix\url{http://link.aps.org/doi/10.1103/PhysRevLett.109.263004}

\bibitem{Sherman2012}
J.~A. Sherman, N.~D. Lemke, N.~Hinkley, M.~Pizzocaro, R.~W. Fox, A.~D. Ludlow,
  C.~W. Oates,
  \href{http://link.aps.org/doi/10.1103/PhysRevLett.108.153002}{High-accuracy
  measurement of atomic polarizability in an optical lattice clock}, Phys. Rev.
  Lett. 108 (2012) 153002.
\newblock \href {http://dx.doi.org/10.1103/PhysRevLett.108.153002}
  {\path{doi:10.1103/PhysRevLett.108.153002}}.
\newline\urlprefix\url{http://link.aps.org/doi/10.1103/PhysRevLett.108.153002}

\bibitem{Dolezal2015}
M.~Dole\v{z}al, P.~Balling, P.~B.~R. Nisbet-Jones, S.~A. King, J.~M. Jones,
  H.~A. Klein, P.~Gill, T.~Lindvall, A.~E. Wallin, M.~Merimaa, C.~Tamm,
  C.~Sanner, N.~Huntemann, N.~Scharnhorst, I.~D. Leroux, P.~O. Schmidt,
  T.~Burgermeister, T.~E. Mehlst\"{a}ubler, E.~Peik,
  \href{http://stacks.iop.org/0026-1394/52/i=6/a=842}{Analysis of thermal
  radiation in ion traps for optical frequency standards}, Metrologia 52~(6)
  (2015) 842.
\newline\urlprefix\url{http://stacks.iop.org/0026-1394/52/i=6/a=842}

\bibitem{Lodewyck2012}
J.~Lodewyck, M.~Zawada, L.~Lorini, M.~Gurov, P.~Lemonde, Observation and
  cancellation of a perturbing dc stark shift in strontium optical lattice
  clocks, Ultrasonics, Ferroelectrics and Frequency Control, IEEE Transactions
  on 59~(3) (2012) 411 --415.
\newblock \href {http://dx.doi.org/10.1109/TUFFC.2012.2209}
  {\path{doi:10.1109/TUFFC.2012.2209}}.

\bibitem{Westergaard2011}
P.~G. Westergaard, J.~Lodewyck, L.~Lorini, A.~Lecallier, E.~A. Burt, M.~Zawada,
  J.~Millo, P.~Lemonde, Lattice-induced frequency shifts in {S}r optical
  lattice clocks at the $10^{-17}$ level, Phys. Rev. Lett. 106~(21) (2011)
  210801.
\newblock \href {http://dx.doi.org/10.1103/PhysRevLett.106.210801}
  {\path{doi:10.1103/PhysRevLett.106.210801}}.

\bibitem{Brown2017a}
R.~C. Brown, N.~B. Phillips, K.~Beloy, W.~F. McGrew, M.~Schioppo, R.~J. Fasano,
  G.~Milani, X.~Zhang, N.~Hinkley, H.~Leopardi, T.~H. Yoon, D.~Nicolodi, T.~M.
  Fortier, A.~D. Ludlow, Hyperpolarizability and {{Operational Magic
  Wavelength}} in an {{Optical Lattice Clock}}, Phys. Rev. Lett. 119~(25)
  (2017) 253001.
\newblock \href {http://dx.doi.org/10.1103/PhysRevLett.119.253001}
  {\path{doi:10.1103/PhysRevLett.119.253001}}.

\bibitem{Campbell2017a}
S.~L. Campbell, R.~B. Hutson, G.~E. Marti, A.~Goban, N.~Darkwah~Oppong, R.~L.
  McNally, L.~Sonderhouse, J.~M. Robinson, W.~Zhang, B.~J. Bloom, J.~Ye,
  \href{http://science.sciencemag.org/content/358/6359/90}{A fermi-degenerate
  three-dimensional optical lattice clock}, Science 358~(6359) (2017) 90--94.
\newblock \href
  {http://arxiv.org/abs/http://science.sciencemag.org/content/358/6359/90.full.pdf}
  {\path{arXiv:http://science.sciencemag.org/content/358/6359/90.full.pdf}},
  \href {http://dx.doi.org/10.1126/science.aam5538}
  {\path{doi:10.1126/science.aam5538}}.
\newline\urlprefix\url{http://science.sciencemag.org/content/358/6359/90}

\bibitem{Yudin2010}
V.~I. Yudin, A.~V. Taichenachev, C.~W. Oates, Z.~W. Barber, N.~D. Lemke, A.~D.
  Ludlow, U.~Sterr, C.~Lisdat, F.~Riehle, Hyper-ramsey spectroscopy of optical
  clock transitions, Phys. Rev. A 82~(1) (2010) 011804.
\newblock \href {http://dx.doi.org/10.1103/PhysRevA.82.011804}
  {\path{doi:10.1103/PhysRevA.82.011804}}.

\bibitem{Chou2010}
C.~W. Chou, D.~B. Hume, J.~C.~J. Koelemeij, D.~J. Wineland, T.~Rosenband,
  Frequency comparison of two high-accuracy {A}l$^{+}$ optical clocks, Phys.
  Rev. Lett. 104~(7) (2010) 070802.
\newblock \href {http://dx.doi.org/10.1103/PhysRevLett.104.070802}
  {\path{doi:10.1103/PhysRevLett.104.070802}}.

\bibitem{LeTargat2013a}
R.~Le~Targat, L.~Lorini, Y.~Le~Coq, M.~Zawada, J.~Gu{\'e}na, M.~Abgrall,
  M.~Gurov, P.~Rosenbusch, D.~G. Rovera, B.~Nag{\'o}rny, R.~Gartman, P.~G.
  Westergaard, M.~E. Tobar, M.~Lours, G.~Santarelli, A.~Clairon, S.~Bize,
  P.~Laurent, P.~Lemonde, J.~Lodewyck,
  \href{http://dx.doi.org/10.1038/ncomms3109}{Experimental realization of an
  optical second with strontium lattice clocks}, Nat Commun 4 (2013) 2109.
\newline\urlprefix\url{http://dx.doi.org/10.1038/ncomms3109}

\bibitem{Bloom2013}
J.~R. W. S. L. C. M. B. X. Z. W. Z. S. L. B. J.~Y. B.~J.~Bloom, T.
  L.~Nicholson, A new generation of atomic clocks: Accuracy and stability at
  the $10^{-18}$ level, arXiv:1309.1137.

\bibitem{Huang2016}
Y.~Huang, H.~Guan, P.~Liu, W.~Bian, L.~Ma, K.~Liang, T.~Li, K.~Gao,
  \href{http://link.aps.org/doi/10.1103/PhysRevLett.116.013001}{Frequency
  comparison of two $^{40}${Ca}$^{+}$ optical clocks with an uncertainty at the
  ${10}^{-17}$ level}, Phys. Rev. Lett. 116 (2016) 013001.
\newblock \href {http://dx.doi.org/10.1103/PhysRevLett.116.013001}
  {\path{doi:10.1103/PhysRevLett.116.013001}}.
\newline\urlprefix\url{http://link.aps.org/doi/10.1103/PhysRevLett.116.013001}

\bibitem{Ludlow2015}
A.~D. Ludlow, M.~M. Boyd, J.~Ye, E.~Peik, P.~O. Schmidt,
  \href{http://link.aps.org/doi/10.1103/RevModPhys.87.637}{Optical atomic
  clocks}, Rev. Mod. Phys. 87 (2015) 637--701.
\newblock \href {http://dx.doi.org/10.1103/RevModPhys.87.637}
  {\path{doi:10.1103/RevModPhys.87.637}}.
\newline\urlprefix\url{http://link.aps.org/doi/10.1103/RevModPhys.87.637}

\bibitem{Bergquist1987}
J.~Bergquist, W.~Itano, D.~Wineland, Recoilless optical absorption and doppler
  sidebands of a single trapped ion, Phys. Rev. A 36 (1987) 428.

\bibitem{Paul1990}
W.~Paul, Electromagnetic traps for charged and neutral particles, Rev. Mod.
  Phys. 62 (1990) 531.

\bibitem{Leibfried2003}
D.~Leibfried, R.~Blatt, C.~Monroe, D.~Wineland, Quantum dynamics of single
  trapped ions, Rev. Mod. Phys. 75~(1) (2003) 281--324.
\newblock \href {http://dx.doi.org/10.1103/RevModPhys.75.281}
  {\path{doi:10.1103/RevModPhys.75.281}}.

\bibitem{Katori2001}
H.~Katori, Spectroscopy of strontium atoms in the {L}amb-{D}icke confinement,
  in: Proc. of the 6$^{th}$ Symposium on Frequency Standards and Metrology,
  World scientific, Singapore, 2001, p. 323.

\bibitem{Katori2003a}
H.~Katori, M.~Takamoto, V.~G. Pal\char39{}chikov, V.~D. Ovsiannikov,
  Ultrastable optical clock with neutral atoms in an engineered light shift
  trap, Phys. Rev. Lett. 91~(17) (2003) 173005.
\newblock \href {http://dx.doi.org/10.1103/PhysRevLett.91.173005}
  {\path{doi:10.1103/PhysRevLett.91.173005}}.

\bibitem{Ye2008a}
J.~Ye, H.~J. Kimble, H.~Katori,
  \href{http://www.sciencemag.org/content/320/5884/1734.abstract}{Quantum state
  engineering and precision metrology using state-insensitive light traps},
  Science 320~(5884) (2008) 1734--1738.
\newblock \href
  {http://arxiv.org/abs/http://www.sciencemag.org/content/320/5884/1734.full.pdf}
  {\path{arXiv:http://www.sciencemag.org/content/320/5884/1734.full.pdf}},
  \href {http://dx.doi.org/10.1126/science.1148259}
  {\path{doi:10.1126/science.1148259}}.
\newline\urlprefix\url{http://www.sciencemag.org/content/320/5884/1734.abstract}

\bibitem{Schioppo2017}
S.~M., C.~Brown~R., F.~McGrew~W., H.~N., J.~Fasano~R., B.~K., H.~Yoon~T.,
  M.~G., N.~D., A.~Sherman~J., B.~Phillips~N., W.~Oates~C., D.~Ludlow~A.,
  \href{http://dx.doi.org/10.1038/nphoton.2016.231}{Ultrastable optical clock
  with two cold-atom ensembles}, Nat Photon 11~(1) (2017) 48--52.
\newline\urlprefix\url{http://dx.doi.org/10.1038/nphoton.2016.231}

\bibitem{Hinkley2013a}
N.~Hinkley, J.~A. Sherman, N.~B. Phillips, M.~Schioppo, N.~D. Lemke, K.~Beloy,
  M.~Pizzocaro, C.~W. Oates, A.~D. Ludlow,
  \href{http://www.sciencemag.org/content/341/6151/1215.abstract}{An atomic
  clock with $10^{-18}$ instability}, Science 341~(6151) (2013) 1215--1218.
\newblock \href
  {http://arxiv.org/abs/http://www.sciencemag.org/content/341/6151/1215.full.pdf}
  {\path{arXiv:http://www.sciencemag.org/content/341/6151/1215.full.pdf}},
  \href {http://dx.doi.org/10.1126/science.1240420}
  {\path{doi:10.1126/science.1240420}}.
\newline\urlprefix\url{http://www.sciencemag.org/content/341/6151/1215.abstract}

\bibitem{Takamoto2011}
M.~Takamoto, T.~Takano, H.~Katori,
  \href{http://dx.doi.org/10.1038/nphoton.2011.34}{Frequency comparison of
  optical lattice clocks beyond the {Dick} limit}, Nat Photon advance online
  publication (2011) --.
\newline\urlprefix\url{http://dx.doi.org/10.1038/nphoton.2011.34}

\bibitem{Nicholson2012a}
T.~L. Nicholson, M.~J. Martin, J.~R. Williams, B.~J. Bloom, M.~Bishof, M.~D.
  Swallows, S.~L. Campbell, J.~Ye,
  \href{http://link.aps.org/doi/10.1103/PhysRevLett.109.230801}{Comparison of
  two independent sr optical clocks with
  $1\mathbf{\ifmmode\times\else\texttimes\fi{}}{10}^{-17}$ stability at
  ${10}^{3}$~s}, Phys. Rev. Lett. 109 (2012) 230801.
\newblock \href {http://dx.doi.org/10.1103/PhysRevLett.109.230801}
  {\path{doi:10.1103/PhysRevLett.109.230801}}.
\newline\urlprefix\url{http://link.aps.org/doi/10.1103/PhysRevLett.109.230801}

\bibitem{AlMasoudi2015}
A.~Al-Masoudi, S.~D\"orscher, S.~H\"afner, U.~Sterr, C.~Lisdat,
  \href{http://link.aps.org/doi/10.1103/PhysRevA.92.063814}{Noise and
  instability of an optical lattice clock}, Phys. Rev. A 92 (2015) 063814.
\newblock \href {http://dx.doi.org/10.1103/PhysRevA.92.063814}
  {\path{doi:10.1103/PhysRevA.92.063814}}.
\newline\urlprefix\url{http://link.aps.org/doi/10.1103/PhysRevA.92.063814}

\bibitem{Marti2018}
G.~E. Marti, R.~B. Hutson, A.~Goban, S.~L. Campbell, N.~Poli, J.~Ye, Imaging
  {{Optical Frequencies}} with 100~$\mu$hz {{Precision}} and 1.1~$\mu$m
  {{Resolution}}, Phys. Rev. Lett. 120~(10) (2018) 103201.
\newblock \href {http://dx.doi.org/10.1103/PhysRevLett.120.103201}
  {\path{doi:10.1103/PhysRevLett.120.103201}}.

\bibitem{Chou2011a}
C.~W. Chou, D.~B. Hume, M.~J. Thorpe, D.~J. Wineland, T.~Rosenband, Quantum
  coherence between two atoms beyond ${Q}=10^{15}$, Phys. Rev. Lett. 106~(16)
  (2011) 160801.
\newblock \href {http://dx.doi.org/10.1103/PhysRevLett.106.160801}
  {\path{doi:10.1103/PhysRevLett.106.160801}}.

\bibitem{Numata2004}
K.~Numata, A.~Kemery, J.~Camp,
  \href{http://link.aps.org/abstract/PRL/v93/e250602}{Thermal-noise limit in
  the frequency stabilization of lasers with rigid cavities}, Phys. Rev. Lett.
  93~(25) (2004) 250602.
\newblock \href {http://dx.doi.org/10.1103/PhysRevLett.93.250602}
  {\path{doi:10.1103/PhysRevLett.93.250602}}.
\newline\urlprefix\url{http://link.aps.org/abstract/PRL/v93/e250602}

\bibitem{Swallows2012}
M.~Swallows, M.~Martin, M.~Bishof, C.~Benko, Y.~Lin, S.~Blatt, A.~Rey, J.~Ye,
  Operating a $^{87}${Sr} optical lattice clock with high precision and at high
  density, Ultrasonics, Ferroelectrics and Frequency Control, IEEE Transactions
  on 59~(3) (2012) 416 --425.
\newblock \href {http://dx.doi.org/10.1109/TUFFC.2012.2210}
  {\path{doi:10.1109/TUFFC.2012.2210}}.

\bibitem{Hafner2015}
S.~H\"{a}fner, S.~Falke, C.~Grebing, S.~Vogt, T.~Legero, M.~Merimaa, C.~Lisdat,
  U.~Sterr, \href{http://ol.osa.org/abstract.cfm?URI=ol-40-9-2112}{$8\times
  10^{-17}$ fractional laser frequency instability with a long room-temperature
  cavity}, Opt. Lett. 40~(9) (2015) 2112--2115.
\newblock \href {http://dx.doi.org/10.1364/OL.40.002112}
  {\path{doi:10.1364/OL.40.002112}}.
\newline\urlprefix\url{http://ol.osa.org/abstract.cfm?URI=ol-40-9-2112}

\bibitem{Kessler2012a}
T.~Kessler, C.~Hagemann, C.~Grebing, T.~Legero, U.~Sterr, F.~Riehle, M.~J.
  Martin, L.~Chen, J.~Ye, A sub-40-{{mHz}}-linewidth laser based on a silicon
  single-crystal optical cavity, Nature Photonics 6~(10) (2012) 687--692.
\newblock \href {http://dx.doi.org/10.1038/nphoton.2012.217}
  {\path{doi:10.1038/nphoton.2012.217}}.

\bibitem{Matei2017}
D.~G. Matei, T.~Legero, S.~H\"afner, C.~Grebing, R.~Weyrich, W.~Zhang,
  L.~Sonderhouse, J.~M. Robinson, J.~Ye, F.~Riehle, U.~Sterr,
  \href{https://link.aps.org/doi/10.1103/PhysRevLett.118.263202}{1.5~$\mu$m
  lasers with sub 10~{mHz} linewidth}, Phys. Rev. Lett. 118 (2017) 263202.
\newblock \href {http://dx.doi.org/10.1103/PhysRevLett.118.263202}
  {\path{doi:10.1103/PhysRevLett.118.263202}}.
\newline\urlprefix\url{https://link.aps.org/doi/10.1103/PhysRevLett.118.263202}

\bibitem{Cole2013a}
G.~D. Cole, W.~Zhang, M.~J. Martin, J.~Ye, M.~Aspelmeyer,
  \href{http://dx.doi.org/10.1038/nphoton.2013.174}{Tenfold reduction of
  brownian noise in high-reflectivity optical coatings}, Nat Photon 7~(8)
  (2013) 644--650.
\newline\urlprefix\url{http://dx.doi.org/10.1038/nphoton.2013.174}

\bibitem{Chen2011d}
Q.-F. Chen, A.~Troshyn, I.~Ernsting, S.~Kayser, S.~Vasilyev, A.~Nevsky,
  S.~Schiller,
  \href{http://link.aps.org/doi/10.1103/PhysRevLett.107.223202}{Spectrally
  narrow, long-term stable optical frequency reference based on a
  {Eu}$^{3+}$:{Y}$_2${SiO}$_5$ crystal at cryogenic temperature}, Phys. Rev.
  Lett. 107 (2011) 223202.
\newblock \href {http://dx.doi.org/10.1103/PhysRevLett.107.223202}
  {\path{doi:10.1103/PhysRevLett.107.223202}}.
\newline\urlprefix\url{http://link.aps.org/doi/10.1103/PhysRevLett.107.223202}

\bibitem{Thorpe2011a}
M.~J. Thorpe, L.~Rippe, T.~M. Fortier, M.~S. Kirchner, T.~Rosenband,
  \href{http://dx.doi.org/10.1038/nphoton.2011.215}{Frequency stabilization to
  $6\times 10^{-16}$ via spectral-hole burning}, Nat Photon 6 (2011) 688.
\newline\urlprefix\url{http://dx.doi.org/10.1038/nphoton.2011.215}

\bibitem{Gobron2017}
O.~Gobron, K.~Jung, N.~Galland, K.~Predehl, R.~L. Targat, A.~Ferrier,
  P.~Goldner, S.~Seidelin, Y.~L. Coq,
  \href{http://www.opticsexpress.org/abstract.cfm?URI=oe-25-13-15539}{Dispersive
  heterodyne probing method for laser frequency stabilization based on spectral
  hole burning in rare-earth doped crystals}, Opt. Express 25~(13) (2017)
  15539--15548.
\newblock \href {http://dx.doi.org/10.1364/OE.25.015539}
  {\path{doi:10.1364/OE.25.015539}}.
\newline\urlprefix\url{http://www.opticsexpress.org/abstract.cfm?URI=oe-25-13-15539}

\bibitem{Yu2007}
D.~Yu, J.~Chen, \href{http://link.aps.org/abstract/PRL/v98/e050801}{Optical
  clock with millihertz linewidth based on a phase-matching effect}, Phys. Rev.
  Lett. 98~(5) (2007) 050801.
\newblock \href {http://dx.doi.org/10.1103/PhysRevLett.98.050801}
  {\path{doi:10.1103/PhysRevLett.98.050801}}.
\newline\urlprefix\url{http://link.aps.org/abstract/PRL/v98/e050801}

\bibitem{Meiser2009a}
D.~Meiser, J.~Ye, D.~R. Carlson, M.~J. Holland,
  \href{http://link.aps.org/abstract/PRL/v102/e163601}{Prospects for a
  millihertz-linewidth laser}, Physical Review Letters 102~(16) (2009) 163601.
\newblock \href {http://dx.doi.org/10.1103/PhysRevLett.102.163601}
  {\path{doi:10.1103/PhysRevLett.102.163601}}.
\newline\urlprefix\url{http://link.aps.org/abstract/PRL/v102/e163601}

\bibitem{Norcia2018}
M.~A. Norcia, J.~R.~K. Cline, J.~A. Muniz, J.~M. Robinson, R.~B. Hutson,
  A.~Goban, G.~E. Marti, J.~Ye, J.~K. Thompson, Frequency {{Measurements}} of
  {{Superradiance}} from the {{Strontium Clock Transition}}, Phys. Rev. X 8~(2)
  (2018) 021036.
\newblock \href {http://dx.doi.org/10.1103/PhysRevX.8.021036}
  {\path{doi:10.1103/PhysRevX.8.021036}}.

\bibitem{Wehner2018}
S.~Wehner, D.~Elkouss, R.~Hanson, Quantum internet: {{A}} vision for the road
  ahead, Science 362~(6412) (2018) eaam9288.
\newblock \href {http://dx.doi.org/10.1126/science.aam9288}
  {\path{doi:10.1126/science.aam9288}}.

\bibitem{Degen2017}
C.~L. Degen, F.~Reinhard, P.~Cappellaro,
  \href{https://link.aps.org/doi/10.1103/RevModPhys.89.035002}{Quantum
  sensing}, Rev. Mod. Phys. 89 (2017) 035002.
\newblock \href {http://dx.doi.org/10.1103/RevModPhys.89.035002}
  {\path{doi:10.1103/RevModPhys.89.035002}}.
\newline\urlprefix\url{https://link.aps.org/doi/10.1103/RevModPhys.89.035002}

\bibitem{Schmidt2005}
P.~O. Schmidt, T.~Rosenband, C.~Langer, W.~M. Itano, J.~C. Bergquist, D.~J.
  Wineland,
  \href{http://www.sciencemag.org/cgi/content/abstract/309/5735/749}{Spectroscopy
  using quantum logic}, Science 309~(5735) (2005) 749--752.
\newline\urlprefix\url{http://www.sciencemag.org/cgi/content/abstract/309/5735/749}

\bibitem{Wineland2013}
D.~J. Wineland, \href{http://link.aps.org/doi/10.1103/RevModPhys.85.1103}{Nobel
  lecture: Superposition, entanglement, and raising {S}chr\"odinger's cat},
  Rev. Mod. Phys. 85 (2013) 1103--1114.
\newblock \href {http://dx.doi.org/10.1103/RevModPhys.85.1103}
  {\path{doi:10.1103/RevModPhys.85.1103}}.
\newline\urlprefix\url{http://link.aps.org/doi/10.1103/RevModPhys.85.1103}

\bibitem{Roos2006}
C.~F. Roos, M.~Chwalla, K.~Kim, M.~Riebe, R.~Blatt,
  \href{http://dx.doi.org/10.1038/nature05101}{``designer atoms'' for quantum
  metrology}, Nature 443~(7109) (2006) 316--319.
\newline\urlprefix\url{http://dx.doi.org/10.1038/nature05101}

\bibitem{Leibfried2002}
D.~Leibfried, B.~DeMarco, V.~Meyer, M.~Rowe, A.~Ben-Kish, J.~Britton, W.~M.
  Itano, B.~Jelenkovi\ifmmode~\acute{c}\else \'{c}\fi{}, C.~Langer,
  T.~Rosenband, D.~J. Wineland,
  \href{http://link.aps.org/doi/10.1103/PhysRevLett.89.247901}{Trapped-ion
  quantum simulator: Experimental application to nonlinear interferometers},
  Phys. Rev. Lett. 89 (2002) 247901.
\newblock \href {http://dx.doi.org/10.1103/PhysRevLett.89.247901}
  {\path{doi:10.1103/PhysRevLett.89.247901}}.
\newline\urlprefix\url{http://link.aps.org/doi/10.1103/PhysRevLett.89.247901}

\bibitem{Keller2019a}
J.~Keller, T.~Burgermeister, D.~Kalincev, A.~Didier, A.~P. Kulosa, T.~Nordmann,
  J.~Kiethe, T.~E. Mehlst\"{a}ubler, Controlling systematic frequency
  uncertainties at the $10^{-19}$ level in linear {Coulomb} crystals, Phys.
  Rev. A 99~(1) (2019) 013405.
\newblock \href {http://dx.doi.org/10.1103/PhysRevA.99.013405}
  {\path{doi:10.1103/PhysRevA.99.013405}}.

\bibitem{Bouchoule2002}
I.~Bouchoule, K.~M{\o}lmer, Preparation of spin-squeezed atomic states by
  optical-phase-shift measurement, Phys. Rev. A 66~(4) (2002) 043811.
\newblock \href {http://dx.doi.org/10.1103/PhysRevA.66.043811}
  {\path{doi:10.1103/PhysRevA.66.043811}}.

\bibitem{Wineland1992}
D.~J. Wineland, J.~J. Bollinger, W.~M. Itano, F.~L. Moore, D.~J. Heinzen,
  \href{http://link.aps.org/doi/10.1103/PhysRevA.46.R6797}{Spin squeezing and
  reduced quantum noise in spectroscopy}, Phys. Rev. A 46 (1992) R6797--R6800.
\newblock \href {http://dx.doi.org/10.1103/PhysRevA.46.R6797}
  {\path{doi:10.1103/PhysRevA.46.R6797}}.
\newline\urlprefix\url{http://link.aps.org/doi/10.1103/PhysRevA.46.R6797}

\bibitem{Windpassinger2008}
P.~J. Windpassinger, D.~Oblak, P.~G. Petrov, M.~Kubasik, M.~Saffman, C.~L.~G.
  Alzar, J.~Appel, J.~H. M\"{u}ller, N.~Kj{\ae}rgaard, E.~S. Polzik,
  \href{http://link.aps.org/abstract/PRL/v100/e103601}{Nondestructive probing
  of {Rabi} oscillations on the cesium clock transition near the standard
  quantum limit}, Phys. Rev. Lett. 100~(10) (2008) 103601.
\newblock \href {http://dx.doi.org/10.1103/PhysRevLett.100.103601}
  {\path{doi:10.1103/PhysRevLett.100.103601}}.
\newline\urlprefix\url{http://link.aps.org/abstract/PRL/v100/e103601}

\bibitem{Louchet-Chauvet2010}
A.~Louchet-Chauvet, J.~Appel, J.~J. Renema, D.~Oblak, N.~Kjaergaard, E.~S.
  Polzik,
  \href{http://stacks.iop.org/1367-2630/12/i=6/a=065032}{Entanglement-assisted
  atomic clock beyond the projection noise limit}, New Journal of Physics
  12~(6) (2010) 065032.
\newline\urlprefix\url{http://stacks.iop.org/1367-2630/12/i=6/a=065032}

\bibitem{Hosten2016}
O.~Hosten, N.~J. Engelsen, R.~Krishnakumar, M.~A. Kasevich,
  \href{http://dx.doi.org/10.1038/nature16176}{Measurement noise 100 times
  lower than the quantum-projection limit using entangled atoms}, Nature
  529~(7587) (2016) 505--508.
\newline\urlprefix\url{http://dx.doi.org/10.1038/nature16176}

\bibitem{Lodewyck2009}
J.~Lodewyck, P.~G. Westergaard, P.~Lemonde,
  \href{http://link.aps.org/abstract/PRA/v79/e061401}{Nondestructive
  measurement of the transition probability in a {Sr} optical lattice clock},
  Physical Review A (Atomic, Molecular, and Optical Physics) 79~(6) (2009)
  061401.
\newblock \href {http://dx.doi.org/10.1103/PhysRevA.79.061401}
  {\path{doi:10.1103/PhysRevA.79.061401}}.
\newline\urlprefix\url{http://link.aps.org/abstract/PRA/v79/e061401}

\bibitem{Vallet2017a}
G.~Vallet, E.~Bookjans, U.~Eismann, S.~Bilicki, R.~L. Targat, J.~Lodewyck,
  \href{http://stacks.iop.org/1367-2630/19/i=8/a=083002}{A noise-immune
  cavity-assisted non-destructive detection for an optical lattice clock in the
  quantum regime}, New Journal of Physics 19~(8) (2017) 083002.
\newline\urlprefix\url{http://stacks.iop.org/1367-2630/19/i=8/a=083002}

\bibitem{Shiga2012}
N.~Shiga, M.~Takeuchi,
  \href{http://stacks.iop.org/1367-2630/14/i=2/a=023034}{Locking the local
  oscillator phase to the atomic phase via weak measurement}, New Journal of
  Physics 14~(2) (2012) 023034.
\newline\urlprefix\url{http://stacks.iop.org/1367-2630/14/i=2/a=023034}

\bibitem{Kohlhaas2015}
R.~Kohlhaas, A.~Bertoldi, E.~Cantin, A.~Aspect, A.~Landragin, P.~Bouyer,
  \href{http://link.aps.org/doi/10.1103/PhysRevX.5.021011}{Phase locking a
  clock oscillator to a coherent atomic ensemble}, Phys. Rev. X 5 (2015)
  021011.
\newblock \href {http://dx.doi.org/10.1103/PhysRevX.5.021011}
  {\path{doi:10.1103/PhysRevX.5.021011}}.
\newline\urlprefix\url{http://link.aps.org/doi/10.1103/PhysRevX.5.021011}

\bibitem{Kruse2016}
I.~Kruse, K.~Lange, J.~Peise, B.~Lücke, L.~Pezzè, J.~Arlt, W.~Ertmer,
  C.~Lisdat, L.~Santos, A.~Smerzi, C.~Klempt, Improvement of an {{Atomic
  Clock}} using {{Squeezed Vacuum}}, Phys. Rev. Lett. 117~(14) (2016) 143004.
\newblock \href {http://dx.doi.org/10.1103/PhysRevLett.117.143004}
  {\path{doi:10.1103/PhysRevLett.117.143004}}.

\bibitem{Leroux2010a}
I.~D. Leroux, M.~H. Schleier-Smith, V.~Vuleti\ifmmode~\acute{c}\else
  \'{c}\fi{}, Orientation-dependent entanglement lifetime in a squeezed atomic
  clock, Phys. Rev. Lett. 104~(25) (2010) 250801.
\newblock \href {http://dx.doi.org/10.1103/PhysRevLett.104.250801}
  {\path{doi:10.1103/PhysRevLett.104.250801}}.

\bibitem{Leroux2010}
I.~D. Leroux, M.~H. Schleier-Smith, V.~Vuleti\ifmmode~\acute{c}\else
  \'{c}\fi{}, Implementation of cavity squeezing of a collective atomic spin,
  Phys. Rev. Lett. 104~(7) (2010) 073602.
\newblock \href {http://dx.doi.org/10.1103/PhysRevLett.104.073602}
  {\path{doi:10.1103/PhysRevLett.104.073602}}.

\bibitem{Diddams2010}
S.~A. Diddams, \href{http://josab.osa.org/abstract.cfm?URI=josab-27-11-B51}{The
  evolving optical frequency comb $[$invited$]$}, J. Opt. Soc. Am. B 27~(11)
  (2010) B51--B62.
\newblock \href {http://dx.doi.org/10.1364/JOSAB.27.000B51}
  {\path{doi:10.1364/JOSAB.27.000B51}}.
\newline\urlprefix\url{http://josab.osa.org/abstract.cfm?URI=josab-27-11-B51}

\bibitem{Newbury2011}
N.~R. Newbury, Searching for applications with a fine-tooth comb, Nature
  Photonics 5 (2011) 186--188.
\newblock \href {http://dx.doi.org/10.1038/nphoton.2011.38}
  {\path{doi:10.1038/nphoton.2011.38}}.

\bibitem{Nicolodi2014}
D.~Nicolodi, B.~Argence, W.~Zhang, R.~Le~Targat, G.~Santarelli, Y.~Le~Coq,
  \href{http://dx.doi.org/10.1038/nphoton.2013.361}{Spectral purity transfer
  between optical wavelengths at the $10^{-18}$ level}, Nat Photon 8 (2014)
  219.
\newline\urlprefix\url{http://dx.doi.org/10.1038/nphoton.2013.361}

\bibitem{Xie2017a}
X.~Xie, R.~Bouchand, D.~Nicolodi, M.~Giunta, W.~Hänsel, M.~Lezius, A.~Joshi,
  S.~Datta, C.~Alexandre, M.~Lours, P.-A. Tremblin, G.~Santarelli,
  R.~Holzwarth, Y.~Le~Coq,
  \href{http://dx.doi.org/10.1038/nphoton.2016.215}{Photonic microwave signals
  with zeptosecond-level absolute timing noise}, Nat Photon 11~(1) (2017)
  44--47.
\newline\urlprefix\url{http://dx.doi.org/10.1038/nphoton.2016.215}

\bibitem{Baynes2015}
F.~N. Baynes, F.~Quinlan, T.~M. Fortier, Q.~Zhou, A.~Beling, J.~C. Campbell,
  S.~A. Diddams,
  \href{http://www.opticsinfobase.org/optica/abstract.cfm?URI=optica-2-2-141}{Attosecond
  timing in optical-to-electrical conversion}, Optica 2~(2) (2015) 141--146.
\newblock \href {http://dx.doi.org/10.1364/OPTICA.2.000141}
  {\path{doi:10.1364/OPTICA.2.000141}}.
\newline\urlprefix\url{http://www.opticsinfobase.org/optica/abstract.cfm?URI=optica-2-2-141}

\bibitem{Ohmae2017}
N.~Ohmae, N.~Kuse, M.~E. Fermann, H.~Katori,
  \href{http://stacks.iop.org/1882-0786/10/i=6/a=062503}{All-polarization-maintaining,
  single-port er:fiber comb for high-stability comparison of optical lattice
  clocks}, Applied Physics Express 10~(6) (2017) 062503.
\newline\urlprefix\url{http://stacks.iop.org/1882-0786/10/i=6/a=062503}

\bibitem{LeCoq2012}
Y.~{Le Coq}, {et al.}, Peignes de fr\'equences femtosecondes pour la mesure des
  fr\'equences optiques, Revue Française de M{\'e}trologie 32 (2012) 35.

\bibitem{Millo2009a}
J.~Millo, M.~Abgrall, M.~Lours, E.~M.~L. English, H.~Jiang, J.~Gu{\'e}na,
  A.~Clairon, M.~E. Tobar, S.~Bize, Y.~L. Coq, G.~Santarelli,
  \href{http://link.aip.org/link/?APL/94/141105/1}{Ultralow noise microwave
  generation with fiber-based optical frequency comb and application to atomic
  fountain clock}, Applied Physics Letters 94~(14) (2009) 141105.
\newblock \href {http://dx.doi.org/10.1063/1.3112574}
  {\path{doi:10.1063/1.3112574}}.
\newline\urlprefix\url{http://link.aip.org/link/?APL/94/141105/1}

\bibitem{Weyers2009a}
S.~Weyers, B.~Lipphardt, H.~Schnatz,
  \href{http://link.aps.org/abstract/PRA/v79/e031803}{Reaching the quantum
  limit in a fountain clock using a microwave oscillator phase locked to an
  ultrastable laser}, Phys. Rev. A 79~(3) (2009) 031803.
\newblock \href {http://dx.doi.org/10.1103/PhysRevA.79.031803}
  {\path{doi:10.1103/PhysRevA.79.031803}}.
\newline\urlprefix\url{http://link.aps.org/abstract/PRA/v79/e031803}

\bibitem{Lipphardt2017}
B.~Lipphardt, V.~Gerginov, S.~Weyers, Optical stabilization of a microwave
  oscillator for fountain clock interrogation, IEEE Transactions on
  Ultrasonics, Ferroelectrics, and Frequency Control 64~(4) (2017) 761--766.
\newblock \href {http://dx.doi.org/10.1109/TUFFC.2017.2649044}
  {\path{doi:10.1109/TUFFC.2017.2649044}}.

\bibitem{Lodewyck2016}
J.~Lodewyck, S.~Bilicki, E.~Bookjans, J.-L. Robyr, C.~Shi, G.~Vallet, R.~L.
  Targat, D.~Nicolodi, Y.~L. Coq, J.~Gu\'ena, M.~Abgrall, P.~Rosenbusch,
  S.~Bize, \href{http://stacks.iop.org/0026-1394/53/i=4/a=1123}{Optical to
  microwave clock frequency ratios with a nearly continuous strontium optical
  lattice clock}, Metrologia 53~(4) (2016) 1123.
\newline\urlprefix\url{http://stacks.iop.org/0026-1394/53/i=4/a=1123}

\bibitem{Lisdat2016}
C.~Lisdat, G.~Grosche, N.~Quintin, C.~Shi, S.~Raupach, C.~Grebing, D.~Nicolodi,
  F.~Stefani, A.~Al-Masoudi, S.~Dorscher, S.~Hafner, J.-L. Robyr, N.~Chiodo,
  S.~Bilicki, E.~Bookjans, A.~Koczwara, S.~Koke, A.~Kuhl, F.~Wiotte,
  F.~Meynadier, E.~Camisard, M.~Abgrall, M.~Lours, T.~Legero, H.~Schnatz,
  U.~Sterr, H.~Denker, C.~Chardonnet, Y.~Le~Coq, G.~Santarelli, A.~Amy-Klein,
  R.~Le~Targat, J.~Lodewyck, O.~Lopez, P.-E. Pottie,
  \href{http://dx.doi.org/10.1038/ncomms12443}{A clock network for geodesy and
  fundamental science}, Nat Commun 7 (2016) 12443.
\newline\urlprefix\url{http://dx.doi.org/10.1038/ncomms12443}

\bibitem{Tyumenev2016}
R.~Tyumenev, M.~Favier, S.~Bilicki, E.~Bookjans, R.~L. Targat, J.~Lodewyck,
  D.~Nicolodi, Y.~L. Coq, M.~Abgrall, J.~Guéna, L.~D. Sarlo, S.~Bize,
  \href{http://stacks.iop.org/1367-2630/18/i=11/a=113002}{Comparing a mercury
  optical lattice clock with microwave and optical frequency standards}, New
  Journal of Physics 18~(11) (2016) 113002.
\newline\urlprefix\url{http://stacks.iop.org/1367-2630/18/i=11/a=113002}

\bibitem{Grosche2007}
G.~Grosche, B.~Lipphardt, H.~Schnatz, G.~Santarelli, P.~Lemonde, S.~Bize,
  M.~Lours, F.~Narbonneau, A.~Clairon, O.~Lopez, A.~Amy-Klein, C.~Chardonnet,
  \href{http://www.opticsinfobase.org/abstract.cfm?URI=CLEO-2007-CMKK1}{Transmission
  of an optical carrier frequency over a telecommunication fiber link}, in:
  Conference on Lasers and Electro-Optics/Quantum Electronics and Laser Science
  Conference and Photonic Applications Systems Technologies, Optical Society of
  America, 2007, p. CMKK1.
\newline\urlprefix\url{http://www.opticsinfobase.org/abstract.cfm?URI=CLEO-2007-CMKK1}

\bibitem{Jiang2008}
H.~Jiang, F.~K\'{e}f\'{e}lian, S.~Crane, O.~Lopez, M.~Lours, J.~Millo,
  D.~Holleville, P.~Lemonde, C.~Chardonnet, A.~Amy-Klein, G.~Santarelli,
  \href{http://josab.osa.org/abstract.cfm?URI=josab-25-12-2029}{Long-distance
  frequency transfer over an urban fiber link using optical phase
  stabilization}, J. Opt. Soc. Am. B 25~(12) (2008) 2029--2035.
\newline\urlprefix\url{http://josab.osa.org/abstract.cfm?URI=josab-25-12-2029}

\bibitem{Grosche2009}
G.~Grosche, O.~Terra, K.~Predehl, R.~Holzwarth, B.~Lipphardt, F.~Vogt,
  U.~Sterr, H.~Schnatz,
  \href{http://ol.osa.org/abstract.cfm?URI=ol-34-15-2270}{Optical frequency
  transfer via 146 km fiber link with $10^{-19}$ relative accuracy}, Opt. Lett.
  34~(15) (2009) 2270--2272.
\newblock \href {http://dx.doi.org/10.1364/OL.34.002270}
  {\path{doi:10.1364/OL.34.002270}}.
\newline\urlprefix\url{http://ol.osa.org/abstract.cfm?URI=ol-34-15-2270}

\bibitem{Predehl2012}
K.~Predehl, G.~Grosche, S.~M.~F. Raupach, S.~Droste, O.~Terra, J.~Alnis,
  T.~Legero, T.~W. Hänsch, T.~Udem, R.~Holzwarth, H.~Schnatz,
  \href{http://www.sciencemag.org/content/336/6080/441.abstract}{A
  920-kilometer optical fiber link for frequency metrology at the 19th decimal
  place}, Science 336~(6080) (2012) 441--444.
\newblock \href
  {http://arxiv.org/abs/http://www.sciencemag.org/content/336/6080/441.full.pdf}
  {\path{arXiv:http://www.sciencemag.org/content/336/6080/441.full.pdf}}, \href
  {http://dx.doi.org/10.1126/science.1218442}
  {\path{doi:10.1126/science.1218442}}.
\newline\urlprefix\url{http://www.sciencemag.org/content/336/6080/441.abstract}

\bibitem{Lopez2012a}
O.~Lopez, A.~Haboucha, B.~Chanteau, C.~Chardonnet, A.~Amy-Klein, G.~Santarelli,
  \href{http://www.opticsexpress.org/abstract.cfm?URI=oe-20-21-23518}{Ultra-stable
  long distance optical frequency distribution using the internet fiber
  network}, Opt. Express 20~(21) (2012) 23518--23526.
\newblock \href {http://dx.doi.org/10.1364/OE.20.023518}
  {\path{doi:10.1364/OE.20.023518}}.
\newline\urlprefix\url{http://www.opticsexpress.org/abstract.cfm?URI=oe-20-21-23518}

\bibitem{GuillouCamargo2018}
F.~{Guillou-Camargo}, V.~M\'enoret, E.~Cantin, O.~Lopez, N.~Quintin,
  E.~Camisard, V.~Salmon, J.-M.~L. Merdy, G.~Santarelli, A.~{Amy-Klein}, P.-E.
  Pottie, B.~Desruelle, C.~Chardonnet, First industrial-grade coherent fiber
  link for optical frequency standard dissemination, Appl. Opt., AO 57~(25)
  (2018) 7203--7210.
\newblock \href {http://dx.doi.org/10.1364/AO.57.007203}
  {\path{doi:10.1364/AO.57.007203}}.

\bibitem{Frank2018}
F.~Frank, F.~Stefani, P.~Tuckey, P.~E. Pottie, A {{Sub}}-ps {{Stability Time
  Transfer Method Based}} on {{Optical Modems}}, IEEE Transactions on
  Ultrasonics, Ferroelectrics, and Frequency Control 65~(6) (2018) 1001--1006.
\newblock \href {http://dx.doi.org/10.1109/TUFFC.2018.2833389}
  {\path{doi:10.1109/TUFFC.2018.2833389}}.

\bibitem{Lopez2015}
O.~Lopez, F.~Kéfélian, H.~Jiang, A.~Haboucha, A.~Bercy, F.~Stefani,
  B.~Chanteau, A.~Kanj, D.~Rovera, J.~Achkar, C.~Chardonnet, P.-E. Pottie,
  A.~Amy-Klein, G.~Santarelli,
  \href{http://www.sciencedirect.com/science/article/pii/S1631070515000754}{Frequency
  and time transfer for metrology and beyond using telecommunication network
  fibres}, Comptes Rendus Physique 16~(5) (2015) 531 -- 539, the measurement of
  time / La mesure du temps.
\newblock \href
  {http://dx.doi.org/http://dx.doi.org/10.1016/j.crhy.2015.04.005}
  {\path{doi:http://dx.doi.org/10.1016/j.crhy.2015.04.005}}.
\newline\urlprefix\url{http://www.sciencedirect.com/science/article/pii/S1631070515000754}

\bibitem{Sliwczynski2013}
L.~Sliwczynski, et~al.,
  \href{http://stacks.iop.org/0026-1394/50/i=2/a=133}{Dissemination of time and
  {RF} frequency via a stabilized fibre optic link over a distance of 420~km},
  Metrologia 50~(2) (2013) 133.
\newline\urlprefix\url{http://stacks.iop.org/0026-1394/50/i=2/a=133}

\bibitem{Lopez2012b}
O.~Lopez, A.~Kanj, P.-E. Pottie, D.~Rovera, J.~Achkar, C.~Chardonnet,
  A.~Amy-Klein, G.~Santarelli,
  \href{http://dx.doi.org/10.1007/s00340-012-5241-0}{Simultaneous remote
  transfer of accurate timing and optical frequency over a public fiber
  network}, Applied Physics B: Lasers and Optics 110 (2013) 3,
  10.1007/s00340-012-5241-0.
\newline\urlprefix\url{http://dx.doi.org/10.1007/s00340-012-5241-0}

\bibitem{Gersl2015}
J.~Ger\v{s}l, P.~Delva, P.~Wolf,
  \href{http://stacks.iop.org/0026-1394/52/i=4/a=552}{Relativistic corrections
  for time and frequency transfer in optical fibres}, Metrologia 52~(4) (2015)
  552.
\newline\urlprefix\url{http://stacks.iop.org/0026-1394/52/i=4/a=552}

\bibitem{Schiller2013a}
S.~Schiller,
  \href{http://link.aps.org/doi/10.1103/PhysRevA.87.033823}{Feasibility of
  giant fiber-optic gyroscopes}, Phys. Rev. A 87 (2013) 033823.
\newblock \href {http://dx.doi.org/10.1103/PhysRevA.87.033823}
  {\path{doi:10.1103/PhysRevA.87.033823}}.
\newline\urlprefix\url{http://link.aps.org/doi/10.1103/PhysRevA.87.033823}

\bibitem{Clivati2013}
C.~Clivati, D.~Calonico, G.~A. Costanzo, A.~Mura, M.~Pizzocaro, F.~Levi,
  \href{http://ol.osa.org/abstract.cfm?URI=ol-38-7-1092}{Large-area fiber-optic
  gyroscope on a multiplexed fiber network}, Opt. Lett. 38~(7) (2013)
  1092--1094.
\newblock \href {http://dx.doi.org/10.1364/OL.38.001092}
  {\path{doi:10.1364/OL.38.001092}}.
\newline\urlprefix\url{http://ol.osa.org/abstract.cfm?URI=ol-38-7-1092}

\bibitem{Marra2018}
G.~Marra, C.~Clivati, R.~Luckett, A.~Tampellini, J.~Kronj\"ager, L.~Wright,
  A.~Mura, F.~Levi, S.~Robinson, A.~Xuereb, B.~Baptie, D.~Calonico, Ultrastable
  laser interferometry for earthquake detection with terrestrial and submarine
  cables, Science (2018) eaat4458\href
  {http://dx.doi.org/10.1126/science.aat4458}
  {\path{doi:10.1126/science.aat4458}}.

\bibitem{Delva2017a}
P.~Delva, J.~Lodewyck, S.~Bilicki, E.~Bookjans, G.~Vallet, R.~Le~Targat, P.-E.
  Pottie, C.~Guerlin, F.~Meynadier, C.~Le~Poncin-Lafitte, O.~Lopez,
  A.~Amy-Klein, W.-K. Lee, N.~Quintin, C.~Lisdat, A.~Al-Masoudi, S.~D\"orscher,
  C.~Grebing, G.~Grosche, A.~Kuhl, S.~Raupach, U.~Sterr, I.~R. Hill, R.~Hobson,
  W.~Bowden, J.~Kronj\"ager, G.~Marra, A.~Rolland, F.~N. Baynes, H.~S.
  Margolis, P.~Gill,
  \href{https://link.aps.org/doi/10.1103/PhysRevLett.118.221102}{Test of
  special relativity using a fiber network of optical clocks}, Phys. Rev. Lett.
  118 (2017) 221102.
\newblock \href {http://dx.doi.org/10.1103/PhysRevLett.118.221102}
  {\path{doi:10.1103/PhysRevLett.118.221102}}.
\newline\urlprefix\url{https://link.aps.org/doi/10.1103/PhysRevLett.118.221102}

\bibitem{Reynaud2009}
S.~Reynaud, C.~Salomon, P.~Wolf,
  \href{http://dx.doi.org/10.1007/s11214-009-9539-0}{Testing general relativity
  with atomic clocks}, Space Science Reviews 148~(1) (2009) 233--247.
\newline\urlprefix\url{http://dx.doi.org/10.1007/s11214-009-9539-0}

\bibitem{Mattingly2005}
D.~Mattingly, \href{http://www.livingreviews.org/lrr-2005-5}{Modern tests of
  lorentz invariance}, Living Reviews in Relativity 8~(5).
\newline\urlprefix\url{http://www.livingreviews.org/lrr-2005-5}

\bibitem{Marciano1984}
W.~Marciano, Time variation of the fundamental ``constants'' and {Kaluza-Klein}
  theories, Phys. Rev. Lett. 52 (1984) 489.

\bibitem{Damour1994}
T.~Damour, A.~Polyakov, The string dilaton and a least coupling principle,
  Nucl. Phys. B 423 (1994) 532.

\bibitem{Damour2012}
T.~Damour, \href{http://stacks.iop.org/0264-9381/29/i=18/a=184001}{Theoretical
  aspects of the equivalence principle}, Classical and Quantum Gravity 29~(18)
  (2012) 184001.
\newline\urlprefix\url{http://stacks.iop.org/0264-9381/29/i=18/a=184001}

\bibitem{Guena2012a}
J.~Gu\'ena, M.~Abgrall, D.~Rovera, P.~Rosenbusch, M.~E. Tobar, P.~Laurent,
  A.~Clairon, S.~Bize,
  \href{http://link.aps.org/doi/10.1103/PhysRevLett.109.080801}{Improved tests
  of local position invariance using $^{87}\mathrm{Rb}$ and $^{133}\mathrm{Cs}$
  fountains}, Phys. Rev. Lett. 109 (2012) 080801.
\newblock \href {http://dx.doi.org/10.1103/PhysRevLett.109.080801}
  {\path{doi:10.1103/PhysRevLett.109.080801}}.
\newline\urlprefix\url{http://link.aps.org/doi/10.1103/PhysRevLett.109.080801}

\bibitem{Wcislo2016a}
P.~Wcislo, P.~Morzynski, M.~Bober, A.~Cygan, D.~Lisak, R.~Ciurylo, M.~Zawada,
  \href{http://dx.doi.org/10.1038/s41550-016-0009}{Experimental constraint on
  dark matter detection with optical atomic clocks}, Nature Astronomy 1 (2016)
  0009.
\newline\urlprefix\url{http://dx.doi.org/10.1038/s41550-016-0009}

\bibitem{Delva2018}
P.~Delva, N.~Puchades, E.~Sch\"onemann, F.~Dilssner, C.~Courde, S.~Bertone,
  F.~Gonzalez, A.~Hees, C.~{Le Poncin-Lafitte}, F.~Meynadier,
  R.~{Prieto-Cerdeira}, B.~Sohet, J.~{Ventura-Traveset}, P.~Wolf, Gravitational
  {{Redshift Test Using Eccentric Galileo Satellites}}, Phys. Rev. Lett.
  121~(23) (2018) 231101.
\newblock \href {http://dx.doi.org/10.1103/PhysRevLett.121.231101}
  {\path{doi:10.1103/PhysRevLett.121.231101}}.

\bibitem{Meynadier2018}
F.~Meynadier, P.~Delva, C.~le~{Poncin-Lafitte}, C.~Guerlin, P.~Wolf, Atomic
  clock ensemble in space ({{ACES}}) data analysis, Class. Quantum Grav. 35~(3)
  (2018) 035018.
\newblock \href {http://dx.doi.org/10.1088/1361-6382/aaa279}
  {\path{doi:10.1088/1361-6382/aaa279}}.

\bibitem{Laurent2015}
P.~Laurent, D.~Massonnet, L.~Cacciapuoti, C.~Salomon,
  \href{http://www.sciencedirect.com/science/article/pii/S1631070515000808}{The
  {ACES/PHARAO} space mission}, Comptes Rendus Physique 16~(5) (2015) 540 --
  552, the measurement of time / La mesure du temps.
\newblock \href
  {http://dx.doi.org/http://dx.doi.org/10.1016/j.crhy.2015.05.002}
  {\path{doi:http://dx.doi.org/10.1016/j.crhy.2015.05.002}}.
\newline\urlprefix\url{http://www.sciencedirect.com/science/article/pii/S1631070515000808}

\bibitem{Cacciapuoti2009}
L.~Cacciapuoti, C.~Salomon,
  \href{http://dx.doi.org/10.1140/epjst/e2009-01041-7}{Space clocks and
  fundamental tests: The {ACES} experiment}, The European Physical Journal
  Special Topics 172 (2009) 57--68.
\newblock \href {http://dx.doi.org/10.1140/epjst/e2009-01041-7}
  {\path{doi:10.1140/epjst/e2009-01041-7}}.
\newline\urlprefix\url{http://dx.doi.org/10.1140/epjst/e2009-01041-7}

\bibitem{Cacciapuoti2007}
L.~Cacciapuoti, N.~Dimarcq, G.~Santarelli, P.~Laurent, P.~Lemonde, A.~Clairon,
  P.~Berthoud, A.~Jornod, F.~Reina, S.~Feltham, C.~Salomon,
  \href{http://www.sciencedirect.com/science/article/pii/S0920563206010425}{Atomic
  clock ensemble in space: Scientific objectives and mission status}, Nuclear
  Physics B - Proceedings Supplements 166 (2007) 303 -- 306, proceedings of the
  Third International Conference on Particle and Fundamental Physics in Space.
\newblock \href {http://dx.doi.org/DOI: 10.1016/j.nuclphysbps.2006.12.033}
  {\path{doi:DOI: 10.1016/j.nuclphysbps.2006.12.033}}.
\newline\urlprefix\url{http://www.sciencedirect.com/science/article/pii/S0920563206010425}

\bibitem{Touboul2017}
P.~Touboul, G.~M\'etris, M.~Rodrigues, Y.~Andr\'e, Q.~Baghi, J.~Berg\'e,
  D.~Boulanger, S.~Bremer, P.~Carle, R.~Chhun, B.~Christophe, V.~Cipolla,
  T.~Damour, P.~Danto, H.~Dittus, P.~Fayet, B.~Foulon, C.~Gageant, P.-Y.
  Guidotti, D.~Hagedorn, E.~Hardy, P.-A. Huynh, H.~Inchauspe, P.~Kayser,
  S.~Lala, C.~L\"ammerzahl, V.~Lebat, P.~Leseur, F.~m.~c. Liorzou, M.~List,
  F.~L\"offler, I.~Panet, B.~Pouilloux, P.~Prieur, A.~Rebray, S.~Reynaud,
  B.~Rievers, A.~Robert, H.~Selig, L.~Serron, T.~Sumner, N.~Tanguy, P.~Visser,
  \href{https://link.aps.org/doi/10.1103/PhysRevLett.119.231101}{{MICROSCOPE}
  mission: First results of a space test of the equivalence principle}, Phys.
  Rev. Lett. 119 (2017) 231101.
\newblock \href {http://dx.doi.org/10.1103/PhysRevLett.119.231101}
  {\path{doi:10.1103/PhysRevLett.119.231101}}.
\newline\urlprefix\url{https://link.aps.org/doi/10.1103/PhysRevLett.119.231101}

\bibitem{CCTF2001}
{Consultative Committee for Time and Frequency (CCTF)}, Report of the 15th
  meeting (june 2001) to the international committee for weights and measures,
  Tech. rep., BIPM (2001).

\bibitem{Marion2003}
H.~Marion, F.~{Pereira Dos Santos}, M.~Abgrall, S.~Zhang, Y.~Sortais, S.~Bize,
  I.~Maksimovic, D.~Calonico, J.~Gr\"unert, C.~Mandache, P.~Lemonde,
  G.~Santarelli, P.~Laurent, A.~Clairon, C.~Salomon,
  \href{https://doi.org/10.1103/PhysRevLett.90.150801}{Search for variations of
  fundamental constants using atomic fountain clocks}, Phys. Rev. Lett. 90
  (2003) 150801.
\newblock \href {http://dx.doi.org/10.1103/PhysRevLett.90.150801}
  {\path{doi:10.1103/PhysRevLett.90.150801}}.
\newline\urlprefix\url{https://doi.org/10.1103/PhysRevLett.90.150801}

\bibitem{Bize2004}
S.~Bize, P.~Laurent, M.~Abgrall, H.~Marion, I.~Maksimovic, L.~Cacciapuoti,
  J.~Gr\"{u}nert, C.~Vian, F.~{Pereira dos Santos}, P.~Rosenbusch, P.~Lemonde,
  G.~Santarelli, P.~Wolf, A.~Clairon, A.~Luiten, M.~Tobar, C.~Salomon, Advances
  in $^{133}${C}s fountains, C. R. Physique 5 (2004) 829.

\bibitem{Margolis2015}
H.~S. Margolis, P.~Gill,
  \href{http://stacks.iop.org/0026-1394/52/i=5/a=628}{Least-squares analysis of
  clock frequency comparison data to deduce optimized frequency and frequency
  ratio values}, Metrologia 52~(5) (2015) 628.
\newline\urlprefix\url{http://stacks.iop.org/0026-1394/52/i=5/a=628}

\bibitem{Robertsson2016}
L.~Robertsson, On the evaluation of ultra-high-precision frequency ratio
  measurements: Examining closed loops in a graph theory framework, Metrologia
  53~(6) (2016) 1272--1280.
\newblock \href {http://dx.doi.org/10.1088/0026-1394/53/6/1272}
  {\path{doi:10.1088/0026-1394/53/6/1272}}.

\bibitem{Grebing2016}
C.~Grebing, A.~Al-Masoudi, S.~D\"{o}rscher, S.~H\"{a}fner, V.~Gerginov,
  S.~Weyers, B.~Lipphardt, F.~Riehle, U.~Sterr, C.~Lisdat,
  \href{http://www.osapublishing.org/optica/abstract.cfm?URI=optica-3-6-563}{Realization
  of a timescale with an accurate optical lattice clock}, Optica 3~(6) (2016)
  563--569.
\newblock \href {http://dx.doi.org/10.1364/OPTICA.3.000563}
  {\path{doi:10.1364/OPTICA.3.000563}}.
\newline\urlprefix\url{http://www.osapublishing.org/optica/abstract.cfm?URI=optica-3-6-563}

\bibitem{Rosenband2008}
T.~Rosenband, D.~B. Hume, P.~O. Schmidt, C.~W. Chou, A.~Brusch, L.~Lorini,
  W.~H. Oskay, R.~E. Drullinger, T.~M. Fortier, J.~E. Stalnaker, S.~A. Diddams,
  W.~C. Swann, N.~R. Newbury, W.~M. Itano, D.~J. Wineland, J.~C. Bergquist,
  Frequency ratio of {A}l$^+$ and {H}g$^+$ single-ion optical clocks; metrology
  at the 17th decimal place, Science 319 (2008) 1808.

\bibitem{Nemitz2016a}
N.~Nemitz, T.~Ohkubo, M.~Takamoto, I.~Ushijima, M.~Das, N.~Ohmae, H.~Katori,
  \href{http://dx.doi.org/10.1038/nphoton.2016.20}{Frequency ratio of {Yb} and
  {Sr} clocks with $5\times 10^{-17}$ uncertainty at 150~seconds averaging
  time}, Nat Photon 10~(4) (2016) 258--261.
\newline\urlprefix\url{http://dx.doi.org/10.1038/nphoton.2016.20}

\bibitem{Yamanaka2015a}
K.~Yamanaka, N.~Ohmae, I.~Ushijima, M.~Takamoto, H.~Katori,
  \href{http://link.aps.org/doi/10.1103/PhysRevLett.114.230801}{Frequency ratio
  of $^{199}\mathrm{Hg}$ and $^{87}\mathrm{Sr}$ optical lattice clocks beyond
  the {SI} limit}, Phys. Rev. Lett. 114 (2015) 230801.
\newblock \href {http://dx.doi.org/10.1103/PhysRevLett.114.230801}
  {\path{doi:10.1103/PhysRevLett.114.230801}}.
\newline\urlprefix\url{http://link.aps.org/doi/10.1103/PhysRevLett.114.230801}

\bibitem{Hachisu2018}
H.~Hachisu, F.~Nakagawa, Y.~Hanado, T.~Ido, Months-long real-time generation of
  a time scale based on an optical clock, Scientific Reports 8~(1) (2018) 4243.
\newblock \href {http://dx.doi.org/10.1038/s41598-018-22423-5}
  {\path{doi:10.1038/s41598-018-22423-5}}.

\bibitem{Hachisu2014a}
H.~Hachisu, M.~Fujieda, S.~Nagano, T.~Gotoh, A.~Nogami, T.~Ido, S.~Falke,
  N.~Huntemann, C.~Grebing, B.~Lipphardt, C.~Lisdat, D.~Piester,
  \href{http://ol.osa.org/abstract.cfm?URI=ol-39-14-4072}{Direct comparison of
  optical lattice clocks with an intercontinental baseline of 9000~km}, Opt.
  Lett. 39~(14) (2014) 4072--4075.
\newblock \href {http://dx.doi.org/10.1364/OL.39.004072}
  {\path{doi:10.1364/OL.39.004072}}.
\newline\urlprefix\url{http://ol.osa.org/abstract.cfm?URI=ol-39-14-4072}

\bibitem{Riehle2017}
F.~Riehle, \href{http://dx.doi.org/10.1038/nphoton.2016.235}{Optical clock
  networks}, Nat Photon 11~(1) (2017) 25--31.
\newline\urlprefix\url{http://dx.doi.org/10.1038/nphoton.2016.235}

\bibitem{Herman2018}
D.~Herman, S.~Droste, E.~Baumann, J.~Roslund, D.~Churin, A.~Cingoz, J.-D.
  Desch\^enes, I.~H. Khader, W.~C. Swann, C.~Nelson, N.~R. Newbury,
  I.~Coddington, Femtosecond {{Timekeeping}}: {{Slip}}-{{Free Clockwork}} for
  {{Optical Timescales}}, Phys. Rev. Applied 9~(4) (2018) 044002.
\newblock \href {http://dx.doi.org/10.1103/PhysRevApplied.9.044002}
  {\path{doi:10.1103/PhysRevApplied.9.044002}}.

\bibitem{Delva2019}
P.~Delva, H.~Denker, G.~Lion,
  \href{https://www.springer.com/us/book/9783030114992}{Chronometric geodesy:
  methods and applications}, 2019.
\newline\urlprefix\url{https://www.springer.com/us/book/9783030114992}

\bibitem{Denker2017}
H.~Denker, L.~Timmen, C.~Voigt, S.~Weyers, E.~Peik, H.~S. Margolis, P.~Delva,
  P.~Wolf, G.~Petit, \href{https://doi.org/10.1007/s00190-017-1075-1}{Geodetic
  methods to determine the relativistic redshift at the level of $10^{-18}$ in
  the context of international timescales: a review and practical results},
  Journal of Geodesy\href {http://dx.doi.org/10.1007/s00190-017-1075-1}
  {\path{doi:10.1007/s00190-017-1075-1}}.
\newline\urlprefix\url{https://doi.org/10.1007/s00190-017-1075-1}

\bibitem{Denker2013}
H.~Denker, Regional {{Gravity Field Modeling}}: {{Theory}} and {{Practical
  Results}}, in: G.~Xu (Ed.), Sciences of {{Geodesy}}, Vol. II, Chapter 5,
  {Springer Berlin Heidelberg}, 2013, pp. 185--291.
\newblock \href {http://dx.doi.org/10.1007/978-3-642-28000-9_5}
  {\path{doi:10.1007/978-3-642-28000-9_5}}.

\bibitem{Weiss2017}
M.~Weiss, N.~Pavlis, A re-evaluation of the relativistic redshift on frequency
  standards at {{NIST}}, {{Boulder}}, {{Colorado}}, {{USA}}, Metrologia 54
  (2017) 535.
\newblock \href {http://dx.doi.org/10.1088/1681-7575/aa765c}
  {\path{doi:10.1088/1681-7575/aa765c}}.

\bibitem{Calonico2007}
D.~Calonico, A.~Cina, I.~H. Bendea, F.~Levi, L.~Lorini, A.~Godone,
  \href{http://stacks.iop.org/0026-1394/44/i=5/a=N03}{Gravitational redshift at
  {INRIM}}, Metrologia 44~(5) (2007) L44.
\newline\urlprefix\url{http://stacks.iop.org/0026-1394/44/i=5/a=N03}

\bibitem{Pavlis2003}
N.~K. Pavlis, M.~A. Weiss,
  \href{http://stacks.iop.org/0026-1394/40/i=2/a=311}{The relativistic redshift
  with $3\times^{-17}$ uncertainty at {NIST}, {Boulder}, {Colorado}, {USA}},
  Metrologia 40~(2) (2003) 66.
\newline\urlprefix\url{http://stacks.iop.org/0026-1394/40/i=2/a=311}

\bibitem{Voigt2016}
C.~Voigt, H.~Denker, L.~Timmen, Time-variable gravity potential components for
  optical clock comparisons and the definition of international time scales,
  Metrologia 53~(6) (2016) 1365.
\newblock \href {http://dx.doi.org/10.1088/0026-1394/53/6/1365}
  {\path{doi:10.1088/0026-1394/53/6/1365}}.

\bibitem{Lion2017}
G.~Lion, I.~Panet, P.~Wolf, C.~Guerlin, S.~Bize, P.~Delva,
  \href{http://dx.doi.org/10.1007/s00190-016-0986-6}{Determination of a high
  spatial resolution geopotential model using atomic clock comparisons},
  Journal of Geodesy (2017) 1--15\href
  {http://dx.doi.org/10.1007/s00190-016-0986-6}
  {\path{doi:10.1007/s00190-016-0986-6}}.
\newline\urlprefix\url{http://dx.doi.org/10.1007/s00190-016-0986-6}

\bibitem{Mehlstaubler2018}
T.~E. Mehlst{\"a}ubler, G.~Grosche, C.~Lisdat, P.~O. Schmidt, H.~Denker, Atomic
  clocks for geodesy, Rep. Prog. Phys. 81~(6) (2018) 064401.
\newblock \href {http://dx.doi.org/10.1088/1361-6633/aab409}
  {\path{doi:10.1088/1361-6633/aab409}}.

\bibitem{Muller2018}
J.~M\"uller, D.~Dirkx, S.~M. Kopeikin, G.~Lion, I.~Panet, G.~Petit, P.~N. a.~M.
  Visser, High {{Performance Clocks}} and {{Gravity Field Determination}},
  Space Sci Rev 214~(1) (2018) 5.
\newblock \href {http://dx.doi.org/10.1007/s11214-017-0431-z}
  {\path{doi:10.1007/s11214-017-0431-z}}.

\bibitem{Bondarescu2015a}
R.~Bondarescu, A.~Schärer, A.~Lundgren, G.~Hetényi, N.~Houlié, P.~Jetzer,
  M.~Bondarescu, \href{+ http://dx.doi.org/10.1093/gji/ggv246}{Ground-based
  optical atomic clocks as a tool to monitor vertical surface motion},
  Geophysical Journal International 202~(3) (2015) 1770--1774.
\newblock \href
  {http://arxiv.org/abs//oup/backfile/content_public/journal/gji/202/3/10.1093_gji_ggv246/1/ggv246.pdf}
  {\path{arXiv:/oup/backfile/content_public/journal/gji/202/3/10.1093_gji_ggv246/1/ggv246.pdf}},
  \href {http://dx.doi.org/10.1093/gji/ggv246} {\path{doi:10.1093/gji/ggv246}}.
\newline\urlprefix\url{+ http://dx.doi.org/10.1093/gji/ggv246}

\bibitem{Bondarescu2012}
R.~Bondarescu, M.~Bondarescu, G.~Hetényi, L.~Boschi, P.~Jetzer,
  J.~Balakrishna,
  \href{http://dx.doi.org/10.1111/j.1365-246X.2012.05636.x}{Geophysical
  applicability of atomic clocks: direct continental geoid mapping},
  Geophysical Journal International 191~(1) (2012) 78--82.
\newblock \href {http://dx.doi.org/10.1111/j.1365-246X.2012.05636.x}
  {\path{doi:10.1111/j.1365-246X.2012.05636.x}}.
\newline\urlprefix\url{http://dx.doi.org/10.1111/j.1365-246X.2012.05636.x}

\bibitem{Bongs2015a}
K.~Bongs, Y.~Singh, L.~Smith, W.~He, O.~Kock, D.~Świerad, J.~Hughes,
  S.~Schiller, S.~Alighanbari, S.~Origlia, S.~Vogt, U.~Sterr, C.~Lisdat, R.~L.
  Targat, J.~Lodewyck, D.~Holleville, B.~Venon, S.~Bize, G.~P. Barwood,
  P.~Gill, I.~R. Hill, Y.~B. Ovchinnikov, N.~Poli, G.~M. Tino, J.~Stuhler,
  W.~Kaenders,
  \href{http://www.sciencedirect.com/science/article/pii/S1631070515000602}{Development
  of a strontium optical lattice clock for the {SOC} mission on the {ISS}},
  Comptes Rendus Physique 16~(5) (2015) 553 -- 564, the measurement of time /
  La mesure du temps.
\newblock \href
  {http://dx.doi.org/http://dx.doi.org/10.1016/j.crhy.2015.03.009}
  {\path{doi:http://dx.doi.org/10.1016/j.crhy.2015.03.009}}.
\newline\urlprefix\url{http://www.sciencedirect.com/science/article/pii/S1631070515000602}

\bibitem{Koller2017}
S.~B. Koller, J.~Grotti, S.~Vogt, A.~Al-Masoudi, S.~D\"orscher, S.~H\"afner,
  U.~Sterr, C.~Lisdat,
  \href{http://link.aps.org/doi/10.1103/PhysRevLett.118.073601}{Transportable
  optical lattice clock with $7 \times 10^{-17}$ uncertainty}, Phys. Rev. Lett.
  118 (2017) 073601.
\newblock \href {http://dx.doi.org/10.1103/PhysRevLett.118.073601}
  {\path{doi:10.1103/PhysRevLett.118.073601}}.
\newline\urlprefix\url{http://link.aps.org/doi/10.1103/PhysRevLett.118.073601}

\bibitem{Origlia2018}
S.~Origlia, M.~S. Pramod, S.~Schiller, Y.~Singh, K.~Bongs, R.~Schwarz,
  A.~{Al-Masoudi}, S.~D\"orscher, S.~Herbers, S.~H\"afner, U.~Sterr, C.~Lisdat,
  Towards an optical clock for space: {{Compact}}, high-performance optical
  lattice clock based on bosonic atoms, Phys. Rev. A 98~(5) (2018) 053443.
\newblock \href {http://dx.doi.org/10.1103/PhysRevA.98.053443}
  {\path{doi:10.1103/PhysRevA.98.053443}}.

\bibitem{Grotti2018}
J.~Grotti, S.~Koller, S.~Vogt, S.~H\"afner, U.~Sterr, C.~Lisdat, H.~Denker,
  C.~Voigt, L.~Timmen, A.~Rolland, F.~N. Baynes, H.~S. Margolis, M.~Zampaolo,
  P.~Thoumany, M.~Pizzocaro, B.~Rauf, F.~Bregolin, A.~Tampellini, P.~Barbieri,
  M.~Zucco, G.~A. Costanzo, C.~Clivati, F.~Levi, D.~Calonico, Geodesy and
  metrology with a transportable optical clock, Nature Physics (2018) 1\href
  {http://dx.doi.org/10.1038/s41567-017-0042-3}
  {\path{doi:10.1038/s41567-017-0042-3}}.

\bibitem{Cossel2017}
K.~C. Cossel, E.~M. Waxman, F.~R. Giorgetta, M.~Cermak, I.~R. Coddington,
  D.~Hesselius, S.~Ruben, W.~C. Swann, G.-W. Truong, G.~B. Rieker, N.~R.
  Newbury,
  \href{http://www.osapublishing.org/optica/abstract.cfm?URI=optica-4-7-724}{Open-path
  dual-comb spectroscopy to an airborne retroreflector}, Optica 4~(7) (2017)
  724--728.
\newblock \href {http://dx.doi.org/10.1364/OPTICA.4.000724}
  {\path{doi:10.1364/OPTICA.4.000724}}.
\newline\urlprefix\url{http://www.osapublishing.org/optica/abstract.cfm?URI=optica-4-7-724}

\bibitem{Deschenes2016}
J.-D. Desch\^enes, L.~C. Sinclair, F.~R. Giorgetta, W.~C. Swann, E.~Baumann,
  H.~Bergeron, M.~Cermak, I.~Coddington, N.~R. Newbury,
  \href{http://link.aps.org/doi/10.1103/PhysRevX.6.021016}{Synchronization of
  distant optical clocks at the femtosecond level}, Phys. Rev. X 6 (2016)
  021016.
\newblock \href {http://dx.doi.org/10.1103/PhysRevX.6.021016}
  {\path{doi:10.1103/PhysRevX.6.021016}}.
\newline\urlprefix\url{http://link.aps.org/doi/10.1103/PhysRevX.6.021016}

\bibitem{Giorgetta2013a}
F.~R. Giorgetta, W.~C. Swann, L.~C. Sinclair, E.~Baumann, I.~Coddington, N.~R.
  Newbury, \href{http://dx.doi.org/10.1038/nphoton.2013.69}{Optical two-way
  time and frequency transfer over free space}, Nat Photon 7~(6) (2013)
  434--438.
\newline\urlprefix\url{http://dx.doi.org/10.1038/nphoton.2013.69}

\bibitem{Djerroud2010}
K.~Djerroud, O.~Acef, A.~Clairon, P.~Lemonde, C.~N. Man, E.~Samain, P.~Wolf,
  \href{http://ol.osa.org/abstract.cfm?URI=ol-35-9-1479}{Coherent optical link
  through the turbulent atmosphere}, Opt. Lett. 35~(9) (2010) 1479--1481.
\newline\urlprefix\url{http://ol.osa.org/abstract.cfm?URI=ol-35-9-1479}

\end{thebibliography}

\end{document}